\documentclass[a4paper,11pt]{article}
\usepackage[nointlimits]{amsmath}
\usepackage{amsfonts}
\usepackage{amssymb}
\usepackage{amsmath}
\usepackage{amsthm}
\usepackage{latexsym}
\usepackage{graphicx}
\usepackage{color}
\usepackage{layout}
\usepackage[english]{babel}
\numberwithin{equation}{section} 

\newcommand{\chr}[2]{\displaystyle\genfrac{\{}{\}}{0pt}{}{#1}{#2}}
\def\AA{{\cal A}}
\def\BB{{\cal B}}
\def\CC{{\cal C}}
\def\costante{C}
\def\DD{{\cal D}}
\def\ds{\displaystyle}
\def\EE{{\cal E}}
\def\em{\epsilon_m}

\def\FF{{\cal F}}

\def\HH{{\cal H}}
\def\II{{\cal I}}
\def\inttempo{{\delta}}
\def\LL{{\cal L}}
\def\kA{k_A}
\def\kB{k_B}
\def\kC{k_C}
\def\kD{k_D}
\def\omt{\widetilde{\omega}}
\def\pertx{{F_x}}
\def\perty{{F_y}}
\def\pertz{{F_z}}
\def\primoindice{\lambda}
\def\quartoindice{\rho}
\def\terzoindice{\nu}
\def\torparam{t}
\def\secondoindice{\mu}
\begin{document}


\title{\bf{Constraining spacetime torsion with the Moon and Mercury}}

\author{
Riccardo March
\\
{\small Istituto per le Applicazioni del Calcolo, CNR,
Via dei Taurini 19, 00185 Roma, Italy},
\\
{\small and INFN - Laboratori Nazionali di Frascati (LNF), via E. Fermi 40 Frascati, 00044 Roma,  Italy.}
\\
{\small E-mail: r.march@iac.cnr.it}
\\
\\
Giovanni Bellettini
\\
{\small Dipartimento di Matematica, Universit\`a
di Roma ``Tor Vergata'',}
\\
{\small via della Ricerca Scientifica 1, 00133 Roma, Italy},
\\
{\small and INFN - Laboratori Nazionali di Frascati (LNF),
via E. Fermi 40 Frascati, 00044 Roma, Italy.}
\\
{\small  E-mail:
Giovanni.Bellettini@lnf.infn.it}
\\
\\
Roberto Tauraso
\\
{\small Dipartimento di Matematica, Universit\`a
di Roma ``Tor Vergata'',}
\\
{\small  via della Ricerca Scientifica 1, 00133 Roma, Italy},
\\
{\small and INFN - Laboratori Nazionali di Frascati (LNF),
via E. Fermi 40 Frascati, 00044 Roma,
 Italy.}
\\
{\small E-mail: tauraso@mat.uniroma2.it}
\\
\\
Simone Dell'Agnello
\\
{\small INFN - Laboratori Nazionali di Frascati (LNF),
via E. Fermi 40 Frascati, 00044 Roma, Italy.}
\\
{\small  E-mail: Simone.Dellagnello@lnf.infn.it}
}
\date{}
\maketitle
\thanks{}
\begin{abstract}

We report a search for new gravitational physics phenomena based on
Riemann-Cartan theory of General Relativity including spacetime torsion.
Starting
from the parametrized torsion framework of Mao, Tegmark, Guth and Cabi,
we analyze the motion of test bodies in the presence of torsion,
and in particular we compute the corrections to the perihelion
advance and to the
orbital geodetic precession of a satellite.
We consider the motion of a test body in
a spherically symmetric field,
and the motion of a satellite
in the gravitational field of the Sun and the Earth.
We describe the torsion field by means
of three parameters, and we make use of the autoparallel trajectories,
which in general differ from geodesics when torsion is present.
We derive the specific approximate expression
of the corresponding system of ordinary
differential equations, which are then solved
with methods of Celestial Mechanics.
We calculate the secular
variations of the longitudes of the node and
of the pericenter of the satellite. The computed secular
variations show how the corrections to the perihelion advance
and to the orbital de Sitter effect depend
on the torsion parameters.
All computations are performed under the assumptions
of weak field and slow motion.
To test our predictions, we
use the measurements of  the Moon's geodetic precession from
lunar laser ranging data, and the measurements of
 Mercury's perihelion advance from planetary radar
ranging data.
These measurements are then used to constrain
suitable linear combinations of the torsion parameters.
\end{abstract}

\bigskip


\section{Introduction}\label{sec:introd}
Einstein's theory of General Relativity (GR) successfully describes gravitational
physics in the solar system. Its predictions have passed a wide variey of
precision experimental tests carried out in weak-field and slow-motion regime
with natural bodies and artificial satellites
\cite{Will},
\cite{shapiro-1999}.
These tests include the measurement and verification of
(quoted in terms of relative experimental uncertainty): Mercury's
perihelion advance at 10$^{-3}$ level through planetary radar ranging
\cite{shapiro-mercury};
the redshift of spectral lines of a hydrogen-maser
frequency standard at 10$^{-4}$ level performed with the Gravity Probe A
spacecraft
\cite{Vessot};
the deflection of light by solar gravity via
very-long-baseline (radio) interferometry at 10$^{-4}$ level
\cite{VLBI};
the time-delay by gravitational potential using the Viking spacecrafts
at Mars
\cite{reasenberg-shapiro}
and the Cassini mission at Saturn
(the latter at 10$^{-5}$ level
\cite{Cassini});
the equivalence principle
at 10$^{-13}$ level and the geodetic precession at about six parts in 10$^{-3}$
with lunar laser ranging (LLR) of the Apollo and Lunokhod retroreflectors
\cite{Williams2004};
frame dragging with satellite laser ranging (SLR) of the
LAGEOS satellites and with the Gravity Probe B (GPB) mission (the former at
10$^{-1}$ level
\cite{CiPa:04});
the latter is also expected to yield a very
accurate measurement of the geodetic precession.

The first three measurements above (perihelion precession, gravitational
redshift
and light deflection) are the
three {\it classical tests} originally proposed
by Einstein to verify his theory. The geodetic precession is a relativistic
three-body effect that was predicted \cite{dS:16-2}
by de Sitter in 1916 and observed in 1988
by I.I. Shapiro et al. with an accuracy of 2\% using LLR data
\cite{shap1988}.

In the effort of improving the experimental measurements and
of possibly discovering new physics,
several extensions and
modifications of GR have been developed. One notable attempt
 is the Parametrized Post-Newtonian (PPN) formalism, whose verification
has been the object of continuous and always improving experimental tests. No deviation
from GR has been found so far. However, in presence of new physics  beyond GR,
it is natural to expect and try to measure modifications of the solar system
observables described above, which historically marked the transition from
newtonian physics to relativistic gravity.

In this paper we treat the modification of GR to include spacetime torsion
and motion along autoparallel orbits. Then we use the measured Moon geodetic
precession and Mercury's perihelion advance to test our predictions.
In a companion paper
\cite{LT},
we show how the constraints on torsion provided by the measurement of the geodetic
precession can be used also to constrain spacetime torsion with
the frame dragging experiments on  LAGEOS satellites.

In the near future, improvements
of the limits reported in these two papers may be obtained from the analysis
of Mercury radar ranging (MRR) data taken since 1990
\cite{Will},
from the relentless
analysis of more LLR data from more ground stations, from the mm-level range
precision provided by the new APOLLO station at Apache Point, USA
\cite{Tom}
(operational since 2007) and from the release of the geodetic precession
measurement by the GPB Collaboration.
In the mid-term we expect a
substantial advance from LLR with the deployment of 2$^{nd}$ generation laser
retroreflector payloads with robotic soft-landings on the Moon,
like the missions of the International Lunar Network (ILN), or
similar geophysical network, and like JAXA's Selene-2.
We also expect that in the long term, more stringent limits
can be set with the approved BepiColombo
Mercury orbiter, an ESA Cornerstone mission.

\section{Theoretical framework}\label{sec:theoreticalsetting}
An interesting generalization of GR includes a non-vanishing
torsion. A class of theories allowing the presence of torsion is
based on the extension of Riemann spacetime to Riemann-Cartan spacetime.
The latter has a richer geometric structure, since it is endowed with
a metric $g_{\secondoindice\terzoindice}$ and a connection
$\Gamma^{\primoindice}_{~\secondoindice\terzoindice}$ which is
not the Levi-Civita connection. A compatibility condition between
$g_{\secondoindice\terzoindice}$ and $\Gamma^{\primoindice}_{~\secondoindice\terzoindice}$
is required, namely the covariant derivative of the metric tensor must
vanish identically. Under this assumption the resulting
connection turns out to be non-symmetric, and such a lack of symmetry
gives origin to a non-vanishing torsion tensor. We refer to
\cite{He:76}, \cite{Ha:02}
for the details.

In most torsion theories of gravity, the source of torsion is the
intrinsic spin of matter
\cite{He:76}, \cite{StYa:79}, \cite{YaSt:80}.
A recent review on searches for the role of spin and polarization in gravity can be found in \cite{Wei:10}.
Since the spins of
elementary particles in macroscopic matter are usually oriented
in a random way, such theories predict a negligible amount of torsion
generated by massive bodies. As a consequence spacetime torsion would
be observationally negligible in the solar system.

However, in
\cite{MaTeGuCa:07}
Mao, Tegmark, Guth and Cabi (MTGC)
argue that, if there are theories giving rise to detectable
torsion in the solar system, they should be tested experimentally.
For this reason, in
\cite{MaTeGuCa:07}
a theory-independent framework
based on symmetry arguments is developed, and it is determined by
a set of
parameters describing the torsion
and the metric.
Here, by theory-independent framework, we mean the following:
the metric and the connection are parametrized, around a massive
body, with the help
of symmetry arguments, without reference to a torsion
model based on a specific Lagrangian
(or even on specific field equations).

This
framework can be used to
constrain the above mentioned parameters from
solar system experiments. In particular, MTGC suggest that
GPB is an appropriate experiment for this task.
In
\cite{MaTeGuCa:07}
the authors compute the precession of gyroscopes and find the constraints
that GPB is able to place on the torsion parameters.
In \cite{{HeOb:07}} Hehl and Obukhov argue that measuring torsion requires
intrinsic spin, and
criticize the approach of MTGC, since GPB gyroscopes
do not carry uncompensated elementary particle spin.

MTGC address also the question whether there exists a specific
gravitational Lagrangian which yields a torsion signal detectable
by the GPB experiment. As an example they quote the theory by
Hayashi and Shirafuji (HS) in
\cite{HaSh:79}
 where a massive body
generates a torsion field. In such a theory gravitational forces
are due entirely to spacetime torsion and not to curvature.
The same property is shared by teleparallel theories
\cite{Ko:82}, \cite{MuNi:83}, \cite{Ne:88}, \cite{BlNi:00}, \cite{ArPe:04}.

Then MTGC propose what they call the Einstein-Hayashi-Shirafuji (EHS)
Lagrangian, which is a linear interpolation of
GR and HS Lagrangians.
The main feature of the EHS theory is that it admits both curvature
and torsion.

The EHS model has been criticized by various authors.
Flanagan and Rosenthal show
in \cite{FlRo:07}
that the
linearized EHS theory becomes consistent only if the parameters
in the Lagrangian satisfy suitable relations that, in turn, make
the predictions coinciding with those of GR.
In the paper \cite{PuOb:08}, Puetzfeld and Obukhov derive the equations of motion in the framework
of metric-affine gravity theories, which includes the HS theory, and show
that only test bodies with microstructure (such as spin) can couple to torsion.
The conclusion
is that the EHS theory does not yield a torsion signal detectable for GPB.
For this reason, in
\cite{MaTeGuCa:07}
the EHS Lagrangian is proposed
just as a pedagogical toy model.
In the present paper we will not treat the EHS model.

As also remarked by Flanagan and Rosenthal in \cite{FlRo:07}, the failure
of constructing the specific EHS Lagrangian
does not rule out the possibility that there may exist other torsion
theories
which could be usefully constrained by solar system experiments.
Such torsion models should fit
in the above mentioned theory-independent framework,
similarly to a parametrized post-Newtonian framework including
torsion. We remark that the parametrized formalism of MTGC does not take into account the intrinsic spin of matter
as a possible source of torsion, and in this sense it cannot be a general torsion framework.
However, it is adequate for the
description of torsion around macroscopic massive bodies
in the solar system, such as planets, being the
intrinsic spin  negligible when averaged over such bodies.

For this reason we think it is worthwhile to continue the investigation of
observable effects in the solar system of nonstandard torsion models within
the MTGC parametrized formalism, under suitable working assumptions.
In particular, our aim is
to extend the GPB gyroscopes computations made in \cite{MaTeGuCa:07}
to the case of  motion of planets and satellites.

In the present paper we compute, as an effect
of spacetime torsion,  the corrections
both to the precession of the pericenter of a body orbiting around
a central mass,
and
to the orbital
geodetic precession.
We describe the torsion by means
of three parameters $\torparam_1, \torparam_2, \torparam_3$.
Our computations show that a complete account
of the precessions requires a parametrization of torsion
up to an approximation order higher than the one considered in \cite{MaTeGuCa:07}.

We consider the motion of a test body in
a spherically symmetric
field, and the motion
of a satellite (either the Moon or LAGEOS)
in the gravitational field of the Sun and the Earth.
Since we use a parametrized framework without specifying the coupling of torsion to matter,
we cannot derive the equations of motion of test bodies from the gravitational
field equations. Therefore, in order to compute effects of torsion
on the orbits of planets and satellites, we will work out the implications
of the assumption that the trajectory of a test body is either
an autoparallel curve or a geodesic. Such trajectories do not need to coincide when
torsion is present.
The computations will be carried out under the assumption of weak field and slow motion
of the test body.

We will assume that the motion of the satellite is obtained
by superimposing the fields of the Sun and the Earth, both computed
as if these bodies were at rest. Observe that these assumptions are
satisfied to a sufficient order of approximation in classical General
Relativity.

As in the original paper of de Sitter
\cite{dS:16-2},
we characterize the
motion using the orbital elements of the osculating ellipse.
In terms of these orbital elements, the equations of motion then
reduce to the Lagrangian planetary equations.
We calculate the secular
variations of the longitude $\Omega$ of the node and of the longitude
$\widetilde \omega$ of the pericenter of the satellite. The computed secular
variations show how the corrections to the orbital de Sitter effect depend
on the torsion parameters $\torparam_1, \torparam_2, \torparam_3$.
In addition we calculate the secular variation of the
longitude of the pericenter of a body orbiting around a central mass,
and also in this case we find the corresponding dependence on $t_1$,
$t_2$, $t_3$.
The data from the LLR and MRR measurements are then used to constrain
the relevant linear combinations of the torsion parameters.

Eventually, we consider the geodesic trajectories, and we find that
torsion parameters cannot be constrained by solar system experiments.

The paper is organized as follows.
In Section \ref{sec:spator} we briefly recall the notion of spacetime with
torsion.
In Section \ref{sec:parame} we recall from
\cite{MaTeGuCa:07}
how to
parametrize the metric and torsion tensors under the assumption of
spherical symmetry, and we extend the parametrization up to
a higher order of approximation. In Section \ref{sec:linfullconn} the
connection up to the required order is given.
In Section \ref{sec:autopa} we analyze the equations of autoparallel
trajectories and we derive the related system of ordinary differential
equations to second order. The expression of the system clearly reveals
the perturbation due to torsion with respect to the de Sitter
equations.
In Section \ref{sec:comput} we
calculate the correction due to torsion to the precession
of pericenter.
In Section \ref{sec:cor} we calculate the correction due to torsion
of the third Kepler's law.
In Section \ref{sec:geo}  we investigate the motion of a satellite
in the gravitational field of the Sun and the Earth and we compute
what we can call the perturbative forces due to torsion.
In Section \ref{sec:computorb}
we derive the time evolution of
the orbital elements of the satellite, using the classical perturbation theory
of Celestial Mechanics, particularly the Gauss form of the Lagrange planetary
equations.
In Section \ref{sec:corredue}
we calculate the secular variations
of the orbital elements of the satellite.
In Section \ref{sec:torbia} we give multiplicative torsion biases
relative to the GR predictions.
In Section \ref{sec:limits} we report the constraints on the parameters
of our torsion model from LLR and MRR,
which is one of the main goals of this paper.
In Section \ref{marrazzosifaprete}
we analyze shortly the geodesic trajectories.
In Section \ref{sec:discu}  we summarize the
future prospects of
the LLR and MRR measurements and we discuss
the implications of proposed and approved space missions for the
search reported in this paper.
Eventually, in the Appendix
we confirm using a different formalism
the computation leading to \eqref{erroreguth},
and we show that, in the autoparallel scheme,
torsion produces an effect on the precession of pericenter which was not
taken into account in
\cite{MaTeGuCa:07online}.

\section{Spacetime with torsion}\label{sec:spator}
We briefly recall the basic notions of Riemann-Cartan spacetimes
\cite{He:76}, \cite{Ha:02}.
A spacetime equipped with a Lorentzian metric
$g_{\secondoindice\terzoindice}$ and a  connection
$\Gamma^\primoindice_{~\secondoindice\terzoindice}$
compatible with the metric
is called a Riemann-Cartan spacetime. Compatibility means
that  $\nabla_\secondoindice g_{\terzoindice\primoindice}=0$, where $\nabla$ denotes the
covariant derivative. We recall in particular that
for any vector field $v^\primoindice$
$$
\nabla_\secondoindice v^\primoindice \equiv
\partial_\secondoindice v^\primoindice +
\Gamma^\primoindice_{~\secondoindice\terzoindice} v^\terzoindice.
$$
 The connection is determined
uniquely by $g_{\secondoindice\terzoindice}$ and by the torsion tensor
$$
S_{\secondoindice\terzoindice}^{~~~\primoindice} \equiv \frac{1}{2}\left(
\Gamma_{~\secondoindice\terzoindice}^{\primoindice}  -
\Gamma_{~\terzoindice\secondoindice}^{\primoindice}
\right)
$$
as follows:
\begin{equation}\label{full}
\Gamma^{\primoindice}_{\
\secondoindice\terzoindice}=
\chr{\primoindice}{\secondoindice\terzoindice}
- K_{\secondoindice\terzoindice}^{\ \ \ \primoindice},
\end{equation}
where
 $\{\cdot\}$ is the Levi-Civita connection, and
\begin{equation}\label{contorsione}
K_{\secondoindice\terzoindice}^{\ \ \ \primoindice}
\equiv-S_{\secondoindice\terzoindice}^{\ \ \ \primoindice}-S_{\
\terzoindice\secondoindice}^{\primoindice}-S_{\ \secondoindice\terzoindice}^{\primoindice}
\end{equation}
is the contortion tensor.
In the particular case when
$\Gamma^{\primoindice}_{\
\secondoindice\terzoindice}$ is symmetric with respect to
$\secondoindice,\terzoindice$ the torsion tensor
vanishes. In the present paper we will consider the case
of nonsymmetric connections
$\Gamma^{\primoindice}_{\ \secondoindice\terzoindice}$.

The Riemann tensor of the connection \eqref{full} is given
by
\begin{equation}\label{riemann}
R^{\primoindice}_{~\quartoindice\terzoindice\secondoindice}
=
\partial_{\terzoindice}
\Gamma_{~\secondoindice\quartoindice}^{\primoindice}
-
\partial_{\secondoindice}
\Gamma_{~\terzoindice\quartoindice}^{\primoindice}  +
\Gamma_{~\terzoindice\alpha}^{\primoindice}
\Gamma_{~\secondoindice\quartoindice}^{\alpha}
-
\Gamma_{~\secondoindice\alpha}^{\primoindice}
\Gamma_{~\terzoindice\quartoindice}^{\alpha}.
\end{equation}
The particular case of vanishing torsion tensor corresponds to Riemann
spacetime of GR, while the particular case of vanishing
Riemann tensor corresponds to the Weitzenb\"ock space time
\cite{HaSh:79}.
\section{Parametrizations of metric and torsion in spherical symmetry}\label{sec:parame}
Throughout this paper we use the natural gravitational units where $c=1$ and $G=1$.
In the following we consider a spherically symmetric body
of mass $m$. Introducing spherical coordinates
$(r,\theta,\phi)$,
we
parametrize the metric and torsion
tensors in a region of space (out of the body) where
the dimensionless quantity
$\em\equiv m/r <<1$ (i.e., $r$ is large in comparison
with the Schwarzschild radius of the body). As it will be
shown in the sequel,
such an approximation
is accurate enough for the purpose of our computations\footnote{For example,
considering the field of the Sun of mass $m$ and the Earth and Mercury as test bodies
at a distance $r$ of the order of their orbit semi-major axes, one gets, respectively,
$\em \sim 2 \times 10^{-8}$ and $\em \sim 5 \times 10^{-8}$.}.

We recall that Parametrized Post-Newtonian (PPN) calculations
\cite{Wi:93}
show that a complete account of the
pericenter precession
must involve second order parameters in $\em$ (for instance the PPN parameter
$\beta$). Therefore, assuming
spherical symmetry, we parametrize the metric tensor
$g_{\secondoindice\terzoindice}$ to second order.
Under the assumption of spherical symmetry
the line element has the following general expression in
spherical coordinates:
\begin{equation}\label{themetric}
ds^2 = - h(r) dt^2 + f(r) dr^2 + \alpha(r) r^2 [d
\theta^2 + \sin^2 \theta d\phi^2].
\end{equation}
We can choose $\alpha(r) =1$, and
to second order in $m/r$
we have
\begin{equation}\label{dsFHG}
h(r) = 1+ \HH \frac{m}{r} + \II \frac{m^2}{r^2},
\qquad
f(r) = 1+ \FF \frac{m}{r} + \LL \frac{m^2}{r^2},
\end{equation}
where
$\HH, \FF, \II, \LL$ are
dimensionless parameters.
The metric is then expressed
in nonisotropic spherical coordinates.
In the computations of trajectories
that we will make in the following, only the function $h(r)$
is required to the second order in $\em$, while for
$f(r)$ the first order approximation is sufficient.

We follow the notation of \cite{MaTeGuCa:07} for the
parametrization of the metric tensor: the parametrization \eqref{dsFHG}
reduces to the first order in $\em$ to the one
used in \cite{MaTeGuCa:07}.
In the case of a PPN metric we have
\cite[Section 3.4.1]{CiWh:95}:
%
\begin{equation}\label{paramPPN}
\HH = -2,
\qquad
\FF = 2 \gamma,
\qquad
\II = 2(\beta-\gamma).
\end{equation}
In the present paper all the others
PPN parameters
\cite{Wi:93}
are assumed to be negligible.

When spacetime torsion is present, our calculations
show that a complete account of the precessions must involve a parametrization
of the torsion tensor $S^{~~~\rho}_{\secondoindice\terzoindice}$
 up to second order in $\em$.

We now follow the spherical symmetry arguments used by
\cite{MaTeGuCa:07}:
 to do this, it is convenient
to parametrize the
nonvanishing components of
the torsion tensor
in isotropic
rectangular coordinates $(t, x^1, x^2, x^3)$. We have
\begin{equation}\label{Sfirst}
\begin{aligned}
S_{0i}^{\ \ 0} =& ~ {\mathcal P}(r') \frac{x^i}{(r')^2},
\\
S_{jk}^{\ \ i} =& ~{\mathcal Q}(r') \frac{x^j \delta_{ki} - x^k \delta_{ji}}{(r')^2},
\end{aligned}
\qquad \qquad
i,j,k \in \{1,2,3\},
\end{equation}
where
$r' = \sqrt{(x^1)^2 + (x^2)^2 + (x^3)^2}$,
${\mathcal P}(r'), {\mathcal Q}(r')$
are arbitrary dimensionless functions, and
$\delta_{ij}$ is the Kronecker's symbol.
If we consider such functions as depending on the dimensionless
small quantity $m/r'$, for the purposes of
our computations (see Section \ref{sec:linfullconn}) it is sufficient to
Taylor expand them up to
the second order, hence we write
\begin{equation}\label{krone}
{\mathcal P}(r')
= \torparam_1 \frac{m}{2 r'} + \widetilde \torparam_3 \frac{m^2}{(r')^2}, \qquad
{\mathcal Q}(r') = \torparam_2 \frac{m}{2 r'} + \widetilde \torparam_4 \frac{m^2}{(r')^2},
\end{equation}
where $\torparam_1,\torparam_2, \widetilde \torparam_3, \widetilde
\torparam_4$ are dimensionless constants.
In the particular
case of first order approximation, the above formulas
yield the parametrization used in
\cite{MaTeGuCa:07}.

In order to transform \eqref{Sfirst}
to nonisotropic
spherical coordinates in which the metric \eqref{themetric} is expressed,
it is convenient to first transform to isotropic spherical coordinates
$(r', \theta, \phi)$. We have
for the non vanishing components of the torsion tensor
\begin{equation}\label{Sfirstbis}
\begin{aligned}
S_{tr'}^{\ \ \ t} =&\torparam_1\frac{m}{2(r')^2} + \widetilde
\torparam_3 \frac{m^2}{(r')^3},
\\
S_{r'\theta}^{\ \ \ \theta} =&S_{r'\phi}^{\ \ \ \phi}=
\torparam_2\frac{m}{2(r')^2} +
\widetilde \torparam_4 \frac{m^2}{(r')^3}.
\end{aligned}
\end{equation}
We now further transform \eqref{Sfirstbis} to nonisotropic
spherical coordinates.
To the required second order of accuracy, the transformation
takes the form $(t, r', \theta, \phi) \to (t, r, \theta, \phi)$ with
$r  \simeq r'(1+\frac{\FF}{2} \frac{m}{r'})$. The resulting expression of the
components \eqref{Sfirstbis} of the torsion tensor in such coordinates is:
\begin{equation}\label{Sfirstbissenzatilde}
\begin{aligned}
S_{tr}^{\ \ t} =&\torparam_1\frac{m}{2r^2} + \torparam_3 \frac{m^2}{r^3},
\\
S_{r \theta}^{\ \ \theta} =&S_{r \phi}^{\ \ \phi}=\torparam_2\frac{m}{2r^2} +
\torparam_4 \frac{m^2}{r^3}.
\end{aligned}
\end{equation}
The constants $\torparam_3$ and $\torparam_4$ are related
to $\torparam_1$, $\torparam_2$, $\widetilde \torparam_3$
and $\widetilde \torparam_4$ as follows:
\begin{equation}\label{ttretquattro}
\torparam_3 = \widetilde \torparam_3 -
\frac{\FF}{2} \torparam_1,
\qquad
\torparam_4 = \widetilde \torparam_4 -
\frac{\FF}{2} \torparam_2.
\end{equation}

Therefore
from \eqref{full} and \eqref{contorsione} it follows that
 $\Gamma_{~\secondoindice\terzoindice}^{\primoindice}$ becomes an
explicit  function  of $\torparam_1$, $\torparam_2$, $\torparam_3$, $\torparam_4$,
and the remaining four parameters
involved,
$$
\Gamma^\primoindice_{~\secondoindice
\terzoindice
}= \Gamma^\primoindice_{~\secondoindice
\terzoindice
}\left(
\torparam_1, \torparam_2,
\torparam_3, \torparam_4,\HH, \FF, \II, \LL,
r, \theta,
\phi\right).
$$
Since the metric and the torsion are constructed
so that the compatibility condition $\nabla_\mu g_{\lambda\nu}=0$
is  satisfied, then the metric parameters
are independent of the torsion parameters.
\section{The connection up to second order}\label{sec:linfullconn}
Using \eqref{full} and \eqref{contorsione},
 the nonvanishing components
of the connection in spherical symmetry, approximated to second order
in $\em=m/r$, read as follows:
\begin{eqnarray*}
\Gamma^{t}_{\ tr}
&=& \left(\torparam_1 - \frac{\HH}{2}\right) \frac{m}{r^2} +
\left(\frac{\HH^2}{2} - \II + 2 \torparam_3\right) \frac{m^2}{r^3},
\\
\Gamma^{t}_{\ rt}&=&-\frac{\HH m}{2r^2} + \left(
\frac{\HH^2}{2} - \II
\right)\frac{m^2}{r^3},
\\
\Gamma^{r}_{\ tt}&=&
\left( \torparam_1 - \frac{\HH}{2}\right) \frac{m}{r^2}
+ \left[
\frac{\HH \FF}{2} - \II + \torparam_1(\HH - \FF) + 2
\torparam_3
\right] \frac{m^2}{r^3},
\\
\Gamma^{r}_{\ rr}&=&-
\frac{\FF m}{2 r^2} + \left(\frac{\FF^2}{2} - \LL\right) \frac{m^2}{r^3},
\\
\Gamma^{r}_{\ \theta\theta}&=&
- r + (\FF + \torparam_2) m
- (\FF^2 + \torparam_2 \FF+\LL - 2 \torparam_4) \frac{m^2}{r},
\\
\Gamma^{r}_{\ \phi\phi}&=&
-r\sin^2 \theta+ \left(\FF + \torparam_2\right) m \sin^2 \theta
- \left( \FF^2 + \torparam_2 \FF+\LL - 2\torparam_4\right) \frac{m^2}{r}
\sin^2 \theta,
\\
\Gamma^{\theta}_{\ r\theta}
&=&
\Gamma^{\phi}_{\ r\phi}
= \frac{1}{r},
\\
\Gamma^{\theta}_{\ \theta r}&=&
\Gamma^{\phi}_{\ \phi r}=
\frac{1}{r}-\torparam_2 \frac{m}{r^2}  - 2 \torparam_4 \frac{m^2}{r^3},
\\
\Gamma^{\theta}_{\ \phi\phi}&=&-\sin \theta \cos \theta,
\\
\Gamma^{\phi}_{\ \theta\phi}&=&
\Gamma^{\phi}_{\ \phi\theta}=\frac{\cos \theta }{\sin \theta}.
\end{eqnarray*}
In the computations
of trajectories that we will make in the sequel
only the components
$\Gamma^{r}_{\ tt}$
 $\Gamma^{t}_{\ tr}$,
$\Gamma^{t}_{\ rt}$
 are required to the second
order in $\em$, while for the remaining components the first
order approximation is sufficient. The second order of approximation is
used in  Section \ref{sec:autopa}
for $\Gamma^{r}_{\ tt}$, and in
the Appendix for
$\Gamma^{t}_{\ tr}$,
$\Gamma^{t}_{\ rt}$. It follows that
the parameters $\LL$ and $\torparam_4$
(differently from $\mathcal I$ and $t_3$)
 will not appear
in the next sections, and consequently they will not be considered
in the sequel of the paper.

\section{Equations of autoparallel trajectories}\label{sec:autopa}
The precise form of the equations of motion
of bodies in the gravitational field depends on the way the matter
couples to the metric and the torsion in the Lagrangian (or in the gravitational field equations).
Here we consider the parametrized framework of Section \ref{sec:parame}
without specifying a coupling of torsion to matter, hence without
specifying the field equations.

In a Riemann-Cartan spacetime
there are two different classes of curves, autoparallel and geodesic curves, respectively,
which reduce to the geodesics of Riemann spacetime when torsion is zero \cite{He:76}.
Autoparallels are curves along which
the velocity vector is transported parallel to itself
by the connection $\Gamma^\lambda_{~\mu\nu}$. Geodesics
are curves which are extremals of the length functional, and
along which the velocity vector is transported parallel to itself
by the Levi-Civita connection.
In GR the two types of trajectories
coincide while,
in general, they may differ in the presence of torsion.
They are identical when the torsion is totally skew-symmetric \cite{He:76},
a special condition which is not satisfied within our parametrization.

The equations of motion
of bodies in the gravitational field follow from the field equations, as
a consequence of
the Bianchi identities.
The method of Papapetrou \cite{Pa:51} can be used to derive the equations of motion
of a test body having internal structure, such as for instance a small extended object that may have
either rotational angular momentum or net spin.
In standard torsion theories the trajectories of test bodies with internal structure, in general, are
neither autoparallels nor geodesics \cite{He:76}, \cite{Ha:02},
\cite{BaFr:10}, while
structureless test bodies, such as spinless test particles, follow geodesic trajectories.

In our computations of orbits of either a planet or a satellite
(considered as a test body), we will neglect its internal structure.
In a theory-independent framework we cannot derive the equations of motion from the gravitational field equations.
Therefore we need some working
assumptions on the trajectories of
structureless test bodies: we will investigate the consequences of
the assumption that the trajectories are either autoparallels or geodesics.
Assuming the trajectory to be a geodesic is  natural and consistent with standard torsion theories.
However, we will see in
Section \ref{marrazzosifaprete} that the geodesics
are the same as in the PPN framework. Hence new
predictions related to torsion may arise only when considering
the autoparallel trajectories, which will turn out
to explicitely
depend
on  torsion parameters.
In the following we give some motivations
which make worthwhile the investigation of autoparallel trajectories.

In the paper
\cite{KlPe:99},
Kleinert and Pelster argue that the closure failure of parallelograms
in the presence of torsion adds an additional term to the geodesics
which causes structureless test bodies to follow autoparallel trajectories.
Kleinert et al. also argue in
\cite{KlPe:99}, \cite{KlSh:98} that autoparallel
trajectories are consistent with the principle of inertia.
Hehl and Obukhov in \cite{{HeOb:07}}
criticized
the approach of Kleinert et al.,
since the equations of autoparallel trajectories have not been derived
from the energy-momentum conservation laws. Kleinert investigates this issue in \cite{Kl:00} where
the autoparallel trajectories are derived from the gravitational field equations
via the Bianchi identities, in the case
when  torsion is derived from a scalar potential (see \cite{Ha:02} for a discussion of
such a kind of torsion).

In the papers
\cite{DeTu:02}, \cite{CeDeTu:04},
using a reformulation of the theory
of Brans-Dicke in terms of a connection with torsion \cite{DeTu:82}, Dereli, Tucker et al.
suggest that the autoparallel trajectory of a spinless test particle
is a possibility that has to be taken into account.
In \cite{DeTu:02} the results of the investigation of autoparallel trajectories
are applied to the computation of the orbit of Mercury.
In \cite{DeTu:04a}, \cite{BuDeTu:04}
the equations of autoparallel trajectories
are derived from the gravitational field
equations and Bianchi identities, in the special case
of matter modeled as a pressureless fluid, and
torsion expressed solely in terms of the gradient of the Brans-Dicke
scalar field.

The above mentioned results show that there is an interest
in the autoparallels in spacetime with torsion, which make
worthwhile their investigation in the present paper.
The system of equations of autoparallel trajectories
of a test body reads as
\begin{equation}\label{eq:autopara}
\frac{d^2 x^{\primoindice}}{d\tau^2} +\Gamma^{\primoindice}_{\
\secondoindice\terzoindice}\frac{d
x^{\secondoindice}}{d\tau}\frac{d x^{\terzoindice}}{d\tau}=0,
\end{equation}
where $\tau$ is the proper time \cite{Po:71}.
Notice that only the symmetric part $\frac{1}{2} (\Gamma_{~\mu\nu}^{\lambda}
 + \Gamma_{~\nu\mu}^{\lambda})$ of the connection enters in \eqref{eq:autopara};
in addition,
the totally antisymmetric part of
$S_{\lambda\mu\nu}$ cannot be measured from \eqref{eq:autopara}.

The trajectory of a test body has to be a time-like curve.
Since the connection is compatible with the metric the quantity
$g_{\secondoindice\terzoindice}\frac{dx^{\secondoindice}}{d\tau}\frac{d x^{\terzoindice}}{d\tau}$
is conserved by parallel transport. The tangent vector $\frac{dx^{\secondoindice}}{d\tau}$ to the trajectory
undergoes parallel transport by the connection along the autoparallel. Therefore, an autoparallel
that is time-like at one point has this same orientation everywhere, so that the trajectory is strictly contained in
the light cone determined by $g_{\secondoindice\terzoindice}$, in a neighbourhood of every
of its points. Hence the compatibility of the connection with the metric ensures
that autoparallels fulfil a necessary requirement for causality.

The equations (\ref{eq:autopara}) can be rewritten as
\begin{equation}\label{Weinb}
\frac{d^2 x^{\alpha}}{dt^2} = -\left(
\Gamma^{\alpha}_{\
\secondoindice\terzoindice} - \Gamma^{0}_{~\secondoindice
\terzoindice} \frac{dx^\alpha}{dt}\right) \frac{dx^\secondoindice}{dt}
\frac{dx^\terzoindice}{dt}
\end{equation}
for $\alpha \in \{1,2,3\}$.
In our units
$(\frac{dx^\alpha}{dt})^2$ and
$\frac{d^2x^\alpha}{dt^2}$ are of the order of $\em$.

We use for $x^\alpha$ spherical coordinates $(r,\theta,\phi)$.
Substituting in \eqref{Weinb} the expression of
$\Gamma^\primoindice_{~\secondoindice\terzoindice}$
given in
Section \ref{sec:linfullconn} one gets,
to the
order $\em^2$ of accuracy,
\begin{equation}\label{sistODEsferiche}
\left\{\begin{array}{l}
\ddot{r}
=
\displaystyle
- \frac{m}{r^2}
+
\AA \frac{m}{r^2}
+ \BB
\frac{m^2}{r^3} + {\CC} \frac{m}{r^2} {\dot r}^2
+
\left(
 r +\DD
m
\right) {\dot \theta}^2 + \left(
r + \DD m\right) \sin^2 \theta
~{\dot \phi}^2,
\\
\\
\ddot{\theta} =
\displaystyle \left(
-\frac{2}{r} + \EE \frac{m}{r^2}
\right)
\dot r \dot \theta
+ \sin \theta \cos \theta ~ {\dot \phi}^2,
\\
\\
\ddot{\phi} =
\displaystyle \left(
-\frac{2}{r} +
\EE
\frac{m}{r^2}
\right)
\dot r \dot \phi - 2 \cot \theta ~ \dot \theta ~\dot \phi,
\end{array}
\right.
\end{equation}
where
$$
\begin{cases}
\AA = \displaystyle
-
\torparam_1 + \frac{\HH}{2} + 1,
\\
\\
\BB = \displaystyle -
\frac{\HH \FF}{2}
+ \II - \torparam_1(\HH - \FF) - 2 \torparam_3,
\\
\\
\CC = \torparam_1 - \HH + \displaystyle \frac{\FF}{2},
\\
\\
\DD =
-\FF  - \torparam_2.
\\
\\
\EE = \torparam_1 + \torparam_2 - \HH.
\end{cases}
$$
In order to take into account relativistic corrections,
the right hand sides of \eqref{sistODEsferiche} must be at least
of second order in $\em$;
this is guaranteed if
$\Gamma^{r}_{\ tt}$ is developed
to second order in $\em$, while it is enough to
develop the remaining components
of the connection to the first order.

System \eqref{sistODEsferiche} to lowest order becomes
\begin{equation}\label{mastellasie'dimesso}
\frac{d \vec v}{dt} = - \left(\torparam_1- \frac{\HH}{2}\right) \frac{m}{r^2}
\hat e_r,
\end{equation}
where $\hat e_r$ is the unit vector in the radial direction.
Imposing the Newtonian limit it follows
(see also
\cite[formula (23)]{MaTeGuCa:07})
%
\begin{equation}\label{limiteniutoniano}
\torparam_1 - \frac{\HH}{2}=1,
\end{equation}
hence $\AA =0$.

We now transform \eqref{sistODEsferiche} in rectangular coordinates
$x=r \sin\theta \cos\phi$,
$y=r \sin\theta \sin\phi$,
$z=r \cos\theta$.
Writing
$x^1,x^2,x^3$ for
 $x,y,z$
we get
\begin{equation}\label{hessianappr}
\left\{\begin{array}{l}
\ddot x^\alpha = \displaystyle
- m
\frac{x^\alpha}{r^3}
+ \BB m^2
~\displaystyle
\frac{x^\alpha}{r^4}
+
\EE
m
\displaystyle
\frac{\dot x^\alpha \dot r}{r^2}
+ \DD m
\displaystyle
\frac{x^\alpha}{r^3} \Phi^2 +
\frac{3}{2} \FF m
\displaystyle
\frac{x^\alpha}{r^3} {\dot r}^2, \qquad \alpha = 1,2,3,
\\
\\
\Phi^2 \equiv \displaystyle \sum_{\alpha=1}^3 ({\dot x^\alpha})^2.
\end{array}\right.
\end{equation}
Note that in case of no torsion (i.e., $\torparam_1 = \torparam_2  = \torparam_3=0$) and
when $\FF = 2$ and $\II =0$
 system \eqref{hessianappr} reduces to the
equations of motion of General Relativity in the weak field
approximation.

\section{Precession of pericenter}\label{sec:comput}
From the second equation in \eqref{sistODEsferiche} it follows that
if $\theta$ and $\dot \theta$ vanish at one time then $\theta$ is
identically zero. Therefore, assuming plane motion,
the system \eqref{hessianappr}
can be  written in the form
\begin{equation*}
\left\{\begin{array}{l}
\ddot{x}=
\displaystyle
 -{\frac {m}{{r}^{3}}}x+\pertx,
\\
\\
\ddot{y}=
\displaystyle
 -{\frac {m}{{r}^{3}}}y+\perty,
\end{array}\right.
\end{equation*}
where $(\pertx,\perty)$ is the perturbation with respect to the Newton force,
\begin{equation}\label{exprpert}
\left\{\begin{array}{l}
\pertx=
\displaystyle
\BB
m^2 \frac{x}{r^4}
+
\EE m \frac{\dot x \dot r}{r^2}
+ \DD m \frac{x}{r^3} \Phi^2 + \frac{3}{2} \FF
\displaystyle
m \frac{x}{r^3} {\dot r}^2,
\\
\\
\displaystyle
\perty=
\BB
m^2
\displaystyle
\frac{y}{r^4}
+ \EE
m
\displaystyle
\frac{\dot y \dot r}{r^2}
+
\DD m \frac{y}{r^3} \Phi^2 + \frac{3}{2} \FF
\displaystyle
m \frac{y}{r^3} {\dot r}^2,
\end{array}\right.
\end{equation}
and now $\Phi^2 = \dot x^2 + \dot y^2$.

The vector $(\pertx,\perty)$ can be decomposed in the standard way
along two mutually orthogonal axes  as
\begin{equation}\label{STWkepler}
\left\{\begin{array}{l}
S= \displaystyle
\frac{x}{r} \pertx + \frac{y}{r} \perty,
\\
\\
T=\displaystyle
\frac{\partial (x/r)}{\partial u}
 \pertx
+
\displaystyle
\frac{\partial (y/r)}{\partial u}
\perty.
\end{array}\right.
\end{equation}
Here $S$ is the component along the instantaneous radius vector,
$T$ is the component perpendicular to the instantaneous radius vector in
the direction of motion,
where $u$ is the argument of latitude.
Then, substituting \eqref{exprpert} into \eqref{STWkepler}
gives
\begin{equation}\label{STWreprise}
\left\{\begin{array}{l}
S=\displaystyle
 \BB
\frac{m^2}{r^3}
 + \CC
 m
\displaystyle
\frac{{\dot r}^2}{r^2} + \DD
m {\dot u}^2,
\\
\\
T= \EE
m
\displaystyle
\frac{\dot r \dot u}{r}.
\end{array}
\right.
\end{equation}

Let us now recall
\cite{BrCl:61},  \cite{GeWe:71}
that, using the method of variation of constants,
\begin{equation}\label{ravarconst}
r=\displaystyle
\frac{a(1-e^2)}{1+e\cos v},
\end{equation}
where $a$ is the semimajor axis of the satellite orbit, $e$ is the
eccentricity, $v$ is the true anomaly, and
\begin{equation}\label{rdotvarconst}
\dot{r}=
\displaystyle
\frac{r^2e\sin v}{a(1-e^2)}\dot{v}, \qquad
r^2\dot{v}=\displaystyle
na^2 (1-e^2)^{1/2},
\end{equation}
$n = 2\pi/U$, $U$ the period of revolution. Recall that,
in the Newtonian approximation, $n^2 = m/a^3$ from the Kepler's third law.
Following the standard
astronomical notation, we let $\widetilde \omega$ be the longitude
of the pericenter, and  $u=v+\widetilde \omega$.
We also recall the following planetary equation of Lagrange in
the Gauss form
\cite[Ch. 6, Sec. 6]{GeWe:71}:
%
\begin{equation}\label{gaussform}
\displaystyle
\frac{d\omt}{dt}=
\frac{(1-e^2)^{1/2}}{nae}
\left[
-S\cos v +T\left(1+{r\over
a(1-e^2)}
\right)\sin v
\right].
\end{equation}
Notice that, since
 $S$ and $T$ are of the order $\em^2$, we have that $\frac{d\widetilde \omega}{dt}$
is of the order $\em^{3/2}$. We are
therefore allowed to make the approximation
\begin{equation}\label{udot=vdot}
\dot{u} \simeq \dot{v}.
\end{equation}
Inserting \eqref{STWreprise}
into
\eqref{gaussform},
making also use of \eqref{ravarconst}, \eqref{rdotvarconst},
\eqref{udot=vdot} and the Kepler's third law,
we have
\begin{eqnarray}
\label{preced}
\displaystyle
\frac{d\omt}{dt}&=&
-\frac{\BB m^2}{
n^2 a^4
 e(1-e^2)
} (1 + e \cos v) \cos v \dot v
 - \frac{\CC e m
}{
a(1-e^2)
}  \sin^2 v \cos v \dot v
\\
&&
- \frac{\DD m}{ae (1-e^2)} (1 + e \cos v)^2 \cos v \dot v
+
\frac{
\EE
m }{
a (1-e^2)
}
 \sin^2 v (2 + e \cos v)\dot v.
\end{eqnarray}
According to
perturbation theory,
we now regard the orbital elements on the right
hand side of \eqref{preced} as approximately constants.
Therefore, integrating with respect to $t$, we obtain
\begin{eqnarray}\label{tiinteg}
&& \inttempo \omt
=
-
\frac{\BB m}{a
e (1-e^2)
} \left(
\sin v + \frac{e}{2} v + \frac{e}{4} \sin (2v)
\right)
\nonumber
\\
&& - \frac{\CC e m}{3 a (1-e^2)
} \sin^3 v -
\frac{\DD m}{a e
 (1-e^2)
} \left(
\sin v + e v + \frac{e}{2} \sin (2v) + e^2 \sin v - \frac{e^2}{3}
\sin^3 v
\right)
\nonumber
\\
&& + \frac{\EE m}{a(1-e^2)} \left(
v - \frac{1}{2} \sin(2v) + \frac{e}{3} \sin^3 v
\right).
\end{eqnarray}
%
\subsection{Correction to the precession of pericenter due to
torsion}\label{sec:correz}
Secular terms appear in $\inttempo \widetilde \omega$.
Using the expressions for $\BB$, $\DD$ and $\EE$,
such secular contributions are:
\begin{eqnarray}\label{erroreguth}
(\inttempo \omt)_{\rm sec}
&=&
\displaystyle
\left(
-\HH + \frac{\FF}{2} + 2 \torparam_2 +
\torparam_1^2 - \frac{\II}{2}
+ \torparam_3
\right)
\frac{m}{a(1-e^2)} v.
\end{eqnarray}
If $\torparam_1= \torparam_2 = \torparam_3 =0$, using \eqref{paramPPN}
then we find
\begin{equation}\label{formulaPPN}
(\inttempo \omt)_{\rm sec} = \left(
2 + 2 \gamma - \beta
\right) \frac{m}{a(1-e^2)} v,
\end{equation}
which gives the precession of pericenter in terms of PPN parameters, as
it can be found in
\cite[Chapter 7, formula (7.54)]{Wi:93}.
Moreover, when
$\HH = -2$, $\FF = 2$
and $\II =0$ (i.e., $\beta = \gamma=1$)
 we find the usual expression of
$(\inttempo \omt)_{\rm sec}^{\rm GR}$ given by
General Relativity. In the case of Mercury, such a
precession amounts to $42.98~{\rm arcsec}/{\rm century}$.

Our formula for $\inttempo \widetilde \omega_{{\rm sec}}$ differs
from the formula
\begin{equation}\label{formulamao}
(\inttempo \omt)_{\rm sec}
= \frac{\FF}{2}
(\inttempo \omt)_{\rm sec}^{\rm GR},
\end{equation}
found by Mao et al. in
\cite[formula (C10)]{MaTeGuCa:07}.
Formula \eqref{formulamao}  does not reproduce the PPN
result \eqref{formulaPPN}
when the torsion parameters vanish, though it reproduces
the GR result in the particular case $\HH = -2$ and $\FF = 2$.
In the Appendix we compute the precession of pericenter
following the method used in
\cite{MaTeGuCa:07},
obtaining again
expression \eqref{erroreguth}.

The precession of the pericenter in a reformulation of
the Brans-Dicke theory in terms of a connection with
torsion has also been computed in
\cite{DeTu:02}
by
using autoparallel trajectories.

\section{Correction to Kepler's third law}\label{sec:cor}
In this section we compute the relativistic correction of Kepler's
third law for Earth motion, in the presence of torsion. This result
will be used in the sequel, for the computation of the satellite
geodetic precession. We note that such a correction was
not necessary in the previous computation of the precession
of pericenter at the required order of accuracy.

We introduce the following coordinates.
We take a system of rectangular coordinates centered at the Sun.
The triplet $(\xi, \eta, \zeta)$ denotes the Earth's coordinates in this system, and we
assume that the ecliptic plane coincides
with the plane $\zeta =0$; we assume that
the eccentricity is zero, so that the Earth's orbit is given by
$$
\xi =\rho \cos L, \quad \eta =\rho \sin L,\quad \zeta=0,
$$
where $\rho$ and $L$ denote the radius of the orbit and the longitude of
the Earth, respectively.
Therefore,
system \eqref{sistODEsferiche}, using
$\theta = \pi/2$ and $\dot \rho =0$, yields
\begin{equation}\label{sistODEsfericherho}
-
\frac{m}{\rho^2}
+ \BB
\frac{m^2}{\rho^3}
 + \left(
\rho  + \DD m
\right)
{\dot L}^2 =0,
\end{equation}
where $m$, here and in the sequel, denotes the mass of the Sun, supposed
spherically symmetric.
It follows that $\dot L = \nu_0$, where $\nu_0$  is constant, and that
\begin{equation}\label{eq:msurhosecorder}
\frac{m}{\rho^3} =
\nu_0^2
\frac{1 - \left(
{\FF} + \torparam_2
\right) \frac{m}{\rho}}{\torparam_1
- \frac{\HH}{2} - \BB \frac{m}{\rho}} =
\nu_0^2
\frac{1 - \left(
{\FF} + \torparam_2
\right) \frac{m}{\rho}}{1
 - \BB \frac{m}{\rho}},
\end{equation}
where we have used the Newtonian limit condition \eqref{limiteniutoniano}.
Since the semi-major axis of the Earth's
orbit is large with respect to the
Schwarzschild radius of the Sun, we have
$m/\rho \sim 2 \times 10^{-8} <<1$.
Since in our units $c=1$, we also have that $\nu_0^2 <<1$. It
follows that,
up to second order, equation \eqref{eq:msurhosecorder} becomes
\begin{equation}\label{correctionthirdlaw}
\frac{m}{\rho^3} =
\frac{\nu_0^2}{\torparam_1 - \frac{\HH}{2}}
\left[1-
\frac{
(\torparam_1-\frac{\HH}{2})
(\FF+ \torparam_2)
- \BB
}{\torparam_1 - \frac{\HH}{2}} \frac{m}{\rho}
\right]
=
\nu_0^2
\left[
1-
\left(\FF+ \torparam_2
- \BB\right)
  \frac{m}{\rho}
\right].
\end{equation}
This approximation will be used for the computation of the
satellite geodetic precession in the next section.
\section{Motion of a satellite in the field of the Sun and the Earth}\label{sec:geo}
In this section we investigate the motion of a satellite
(either the Moon or LAGEOS) in the gravitational field
of the Sun and the Earth in presence of torsion.
The coordinates
$(\xi, \eta, \zeta)$ have been defined in Section \ref{sec:cor}.
The triplet $(X,Y,Z)$ denotes the satellite's coordinates, and we set
$$
\Delta^2 \equiv X^2+Y^2+Z^2.
$$
The
satellite's coordinates  with respect to the Sun will be written as
$$
X =\xi+x, \quad Y=\eta+y,\quad Z=\zeta+z,
$$
where $(x,y,z)$ are the coordinates of the satellite with respect to
the Earth. We use the
standard coordinates transformation
\cite{BrCl:61}, \cite{GeWe:71}
used in Celestial Mechanics
\begin{equation}\label{standardcoo}
\left\{\begin{array}{l}
\ds x=  r(\cos u \cos \Omega - \sin u \sin \Omega \cos i),
\\ \\
\ds y= r(\cos u \sin \Omega + \sin u \cos \Omega \cos i),
\\ \\
\ds z= r \sin u \sin i,
\end{array}\right.
\end{equation}
where $r$ is the distance
between the Earth and the satellite,
$u$ is the argument of the latitude,
$\Omega$ is the longitude of the node, and $i$ is the orbital inclination.

We suppose
that the Earth is spherically symmetric.
The semi-major axes of the Moon and LAGEOS orbits around the Earth
are small in comparison with $\rho$, so that in our computations
we will neglect the powers of $r/\rho$ greater than one\footnote{For the Moon and
LAGEOS we have $r/\rho \sim 2.6 \times 10^{-3}$ and $r/\rho \sim 8.5 \times 10^{-3}$ respectively.}.
Hence, also
$m/\Delta <<1$.

We will assume that the motion of the satellite is obtained
by superimposing the fields (i.e., the
 connections as described in Section \ref{sec:linfullconn})
of  the Sun and the Earth, both computed as if these bodies were at rest.
More precisely we assume
\begin{equation}\label{eq:superimpose}
g_{\mu\nu}  =
(g_{\mu\nu})_0 +
(g_{\mu\nu})_1^0
\qquad \qquad
S_{\mu\nu}^{~~~\lambda} =
(S_{\mu\nu}^{~~~\lambda})_0 +
(S_{\mu\nu}^{~~~\lambda})_1^0,
\end{equation}
where
\begin{itemize}
\item[-]
$(g_{\mu\nu})_0$ and
$(S_{\mu\nu}^{~~~\lambda})_0$ are the metric and the torsion
tensors, as given in Section \ref{sec:parame}, taking into account the Sun
only, supposed at rest;
\item[-] $(g_{\mu\nu})_1^0$ and
$(S_{\mu\nu}^{~~~\lambda})_1^0$ are the metric and the torsion tensors taking into
account the Earth only; these tensors are given at each time
by the expressions in Section \ref{sec:parame}, computed as if the
Earth were at rest (at that time).
\end{itemize}

Observe that these assumptions are satisfied to a
sufficient order of approximation
in classical General Relativity
\cite{dS:16-2}, \cite{dS:proc}
if all the terms that give rise to periodic perturbations
are neglected (we are interested only in secular effects).
We also note that
a rigorous justification
of the validity
 of \eqref{eq:superimpose} would probably require
an extension of the parametrized torsion model to the
case of three interacting bodies, which is beyond
the scope of the present paper.

As a consequence we have
$$
\frac{d^2 x^{\alpha}}{dt^2} =
\left(\frac{d^2 x^{\alpha}}{dt^2}\right)_1^0
+
\left(\frac{d^2 X^{\alpha}}{dt^2}\right)_0
-
\left(\frac{d^2 \xi^{\alpha}}{dt^2}\right)_0,
$$
where
we write $x^1,x^2,x^3$ for
$x,y,z$.
Similarly we write $X^1,X^2,X^3$ for
$X,Y,Z$, and
$\xi^1,\xi^2,\xi^3$ for
$\xi,\eta,\zeta$.
Moreover
\begin{itemize}
\item[-]
$\left(\frac{d^2 x^{\alpha}}{dt^2}\right)_1^0$
is the left hand side of \eqref{Weinb}
with the coefficients of the connection computed using
$(g_{\mu\nu})_1^0$ and
$(S_{\mu\nu}^{~~~\lambda})_1^0$~;
\item[-] $\left(\frac{d^2 X^{\alpha}}{dt^2}\right)_0$
is the left hand side of \eqref{Weinb}
with the coefficients computed
using
$(g_{\mu\nu})_0$ and
$(S_{\mu\nu}^{~~~\lambda})_0$~;
\item[-] $\left(\frac{d^2 \xi^{\alpha}}{dt^2}\right)_0$
is the left hand side of \eqref{Weinb}
with the coefficients computed using
$(g_{\mu\nu})_0$ and
$(S_{\mu\nu}^{~~~\lambda})_0$~.
\end{itemize}
The contribution of the term
$\left(\frac{d^2 x^{\alpha}}{dt^2}\right)_1^0$ gives
a secular precession of the perigee which has been computed
in Section \ref{sec:comput}. The other terms represent
the perturbing effect of the Sun. Since all the
perturbations here considered are small enough to allow
us to superimpose them linearly, in what follows
we compute the perturbing effect of the Sun only. With these
assumptions, using the Newtonian limit \eqref{limiteniutoniano}
the right hand members of \eqref{hessianappr} give
\begin{equation}\label{eq:pertsun}
\ddot x^\alpha+
R^\alpha=
\kA A^\alpha
+
\kB B^\alpha
+
\kC C^\alpha
+\kD D^\alpha
 \quad \mbox{for $\alpha=1,2,3$},
\end{equation}
where
\begin{equation*}
\left\{\begin{array}{l}
\ds R^\alpha
=m\left( {X^\alpha \over \Delta^3} -{\xi^\alpha\over \rho^3}\right),
\\
\\
\ds A^\alpha
=m^2\left( {X^\alpha \over \Delta^4} -{\xi^\alpha\over \rho^4}\right),
\\
\\
\ds B^\alpha
=m\left( {\dot\Delta \dot X^\alpha \over \Delta^2} -
{\dot\rho \dot\xi^\alpha\over \rho^2}\right),
\\ \\
\ds C^\alpha
=m\left( {X^\alpha \sum_\sigma (\dot X^\sigma)^2\over \Delta^3} -{\xi^\alpha
 \sum_\sigma (\dot \xi^\sigma)^2\over \rho^3}\right),
\\
\\
\ds
D^\alpha =m\left( {X^\alpha \dot \Delta^2\over \Delta^3} -
{\xi^\alpha \dot\rho^2\over \rho^3}\right),
\end{array}\right.
\end{equation*}
and for notational convenience we set
\begin{equation}\label{eq:KaKbKcKd}
\kA = \BB, \quad \kB = \EE, \quad \kC = \DD,
\quad \kD={3\FF\over 2}.
\end{equation}
The left hand side of \eqref{eq:pertsun} contains the
ordinary Newtonian perturbing function $N^\alpha$, which also
requires a correction $\widehat N^\alpha$, according to the computations in
Section \ref{sec:cor}.
This will be made clear in
 Section \ref{sec:computorb}.

The components $R^\alpha$ and $A^\alpha$, approximated
to the first order
with respect\footnote{From formula
\eqref{ra1e} below expressing $\dot r$, it follows $\vert \dot r\vert \leq
\frac{r \vert \dot v\vert}{1-e}$, so that $\dot r$ is at least small
as $\dot v$, hence $\vert \dot r/\rho\vert$ is smaller than $r/\rho$.
} to $r/\rho$ and $\dot r/\rho$, read
as follows (where we write $R_x$, $R_y$ and $R_z$ for $R^1$, $R^2$
and $R^3$ respectively, and similarly for $A,B$ etc.):
\begin{equation}\label{P1}
\left\{\begin{array}{l}
\ds R_x={m\over \rho^3}
\left(x-3x\cos^2L
\right)
+ P_{R_x},
\\ \\
\ds R_y={m\over \rho^3}\left(y-3y\sin^2L
\right)+ P_{R_y},
\\ \\
\ds R_z={m\over \rho^3}z
+ P_{R_z},
\end{array}\right.
\end{equation}
\begin{equation}\label{P2}
\left\{\begin{array}{l}
\ds A_x=
{m^2\over \rho^4}
\left(x-4x\cos^2L
\right)
+ P_{A_x},
\\ \\
\ds A_y={m^2\over \rho^4}\left(y-4y\sin^2 L \right)
+ P_{A_y},
\\ \\
\ds A_z={m^2\over \rho^4}z + P_{A_z}.
\end{array}\right.
\end{equation}
Here the terms $P_{R^\alpha}$, $P_{A^\alpha}$ denote a finite sum
of addenda with the following property;
each addendum is of the form $f~ (\sin L)^{n_1} (\cos L)^{n_2}$,
where $n_1+n_2$ is odd and $f$ is a factor
independent of $L$. Such terms will give periodic
contributions to the perturbations of the orbital elements:
therefore they will be neglected in the computation of
secular perturbations.

The components $B^\alpha$, $C^\alpha$ and $D^\alpha$, approximated
to the first order with respect to $r/\rho$, $\dot r/\rho$, and
taking into account also the terms in $r \dot r/\rho^2$,
read as follows:
\begin{equation}\label{P3}
\left\{\begin{array}{l}
\ds B_x={m\dot L\over \rho}
(\dot L x-\dot y)\sin^2 L
+ P_{B_x},
\\ \\
\ds
B_y={m\dot L\over \rho}
(\dot L y+\dot x)\cos^2 L
+ P_{B_y},
\\ \\
\ds B_z=
 P_{B_z},
\end{array}\right.
\end{equation}
\begin{equation}\label{P4}
\left\{\begin{array}{l}
\ds C_x
=
{m\dot L\over \rho}\left(
\dot L x
+(-3\dot L x+2\dot y)\cos^2 L
\right)
+ P_{C_x},
\\ \\
\ds C_y
=
{m\dot L\over \rho}\left(\dot L y+(-3\dot L y-2\dot x)\sin^2 L
\right)
+ P_{C_y},
\\ \\
\ds C_z={m\dot L^2\over \rho}z
+ P_{C_z},
\end{array}\right.
\end{equation}
\begin{equation}\label{P5}
\left\{\begin{array}{l}
\ds D_x= P_{D_x},
\\ \\
\ds D_y= P_{D_y},
\\ \\
\ds D_z= P_{D_z}.
\end{array}\right.
\end{equation}
Also here the terms $P_{B^\alpha}$, $P_{C^\alpha}$
and $P_{D^\alpha}$ have the same structure of
$P_{R^\alpha}$ and $P_{A^\alpha}$, thus giving
periodic perturbations of the orbital elements.

\section{Computation of orbital elements via perturbation theory}
\label{sec:computorb}
In this section we introduce the tools from Celestial
Mechanics needed to compute the secular perturbations
of the orbital elements. In the following the orbital elements,
the true anomaly $v$ and the
argument of latitude $u$ will be referred to the satellite's
orbit around Earth.

Using the Newtonian limit condition \eqref{limiteniutoniano} and
the correction to Kepler's third law  \eqref{correctionthirdlaw}
we have
\begin{equation}\label{scompR}
R^\alpha = N^\alpha - k_N \widehat N^\alpha
\qquad {\rm for}~ \alpha=1,2,3,
\end{equation}
where
\begin{equation}\label{nuovomodoN}
\left\{
\begin{array}{l}
N_x =
\nu_0^2 x \left(
1 - 3 \cos^2 L \right)
+ P_{N_x},
\\
\\
N_y =
\nu_0^2 y \left(
1 - 3 \cos^2 L \right)
+ P_{N_y},
\\
\\
N_z =
\nu_0^2 z
+ P_{N_z},
\end{array}
\right.
\end{equation}
\begin{equation}\label{deltaNx}
\left\{
\begin{aligned}
& \widehat N_x =
 \frac{m \nu_0^2}{\rho} x (1 - 3 \cos^2 L)
+ P_{\widehat N_x},
\\
& \widehat N_y =
 \frac{m \nu_0^2}{\rho} y
(1 - 3 \sin^2 L)
+ P_{\widehat N_y},
\\
& \widehat N_z = -
 \frac{m \nu_0^2}{\rho} z
+ P_{\widehat N_z},
\end{aligned}
\right.
\end{equation}
and
\begin{equation}\label{eq:kN}
 k_N = \FF + \torparam_2 - \kA.
\end{equation}
Again the terms $P_{N^\alpha}$ and $P_{\widehat N^\alpha}$
give
periodic contributions to the perturbations of the orbital elements.

In equation \eqref{scompR} $R^\alpha$ is decomposed into the ordinary
Newtonian perturbing function $N^\alpha$ plus a relativistic
correction $- k_N \widehat N^\alpha$, which depends  also
on torsion.

Equations \eqref{eq:pertsun} can be rewritten as
\begin{equation}\label{eq:pertsunnuw}
\ddot x^\alpha+
N^\alpha= F^\alpha
\qquad {\rm for}~ \alpha=1,2,3,
\end{equation}
where $F^\alpha$ is the perturbation with respect to the Newton
force, and it is given by
\begin{equation*}
F^\alpha = k_N \widehat N^\alpha +
\kA A^\alpha
+
\kB B^\alpha
+
\kC C^\alpha
+\kD D^\alpha.
\end{equation*}

Recalling also \eqref{standardcoo}, the perturbation $(\pertx, \perty, \pertz)$
can be decomposed in the standard way along
three mutually orthogonal axes  as
\begin{equation}\label{STW}
\left\{\begin{array}{l}
\ds S=\frac{x}{r} \pertx + \frac{y}{r} \perty + \frac{z}{r} \pertz,
\\
\\
\ds T=\frac{\partial (x/r)}{\partial u} \pertx
+
\frac{\partial (y/r)}{\partial u} \perty
+\frac{\partial (z/r)}{\partial u}
\pertz,
\\
\\
\ds \sin u ~
W=
\frac{\partial (x/r)}{\partial i} \pertx+\frac{\partial (y/r)}{\partial i}
\perty
+\frac{\partial (z/r)}{\partial i} \pertz.
\end{array}\right.
\end{equation}
Here $S$ is the component along the instantaneous radius vector,
$T$ is the component perpendicular to the instantaneous radius vector in
the direction of motion, and $W$ is the component normal to the
osculating plane of the orbit (collinear with the angular momentum vector).

Recalling from Section \ref{sec:cor} that
$\dot L=\nu_0$, $L=\nu_0 t$, we replace
 $W$, $S$, and $T$  with the averages
$$
{\nu_0\over 2\pi}\int_{0}^{2\pi/\nu_0}W\,dt, \qquad
{\nu_0\over 2\pi}\int_{0}^{2\pi/\nu_0}S\,dt, \qquad
{\nu_0\over 2\pi}\int_{0}^{2\pi/\nu_0}T\,dt,
$$
respectively.
Taking these averages has the following consequences:
(i) it eliminates the dependence of $W$
on the trigonometric functions of $L$, hence
the periodic terms in \eqref{P1}-\eqref{P5}
disappear;  (ii) the remaining terms, contributing
to the secular effects,
are multiplied by
a factor depending on $\nu_0$.

In order to compute the components $S$, $T$ and $W$ of the
perturbation, using the method of variation
of constants
\cite{BrCl:61}, \cite{GeWe:71}
we write

\begin{equation*}
\left\{\begin{array}{l}
\ds \dot x=\dot r(\cos u \cos \Omega - \sin u \sin \Omega \cos i)+ r\dot u (-\sin u \cos \Omega - \cos u \sin \Omega \cos i),
\\
\ds \dot y=\dot r(\cos u \sin \Omega + \sin u \cos \Omega \cos i)+ r\dot u (-\sin u \sin \Omega + \cos u \cos \Omega \cos i),
\\
\ds \dot z=\dot r \sin u \sin i +r\cos u \sin i,
\end{array}\right.
\end{equation*}
in order to eliminate $\dot x^\alpha$.

We recall that
 $u=v+\widetilde \omega- \Omega$ (where $\widetilde \omega$
is the longitude of the pericenter), which
allows to make again the approximation
$\dot{u}=\dot{v}$,
and we make use of the area law
$x\dot y-y\dot x=r^2 \dot v \cos i$
in order to simplify
the computations.

We decompose
$$
W = W_A + W_B + W_C + W_{\widehat N},
$$
with obvious meaning of the notation.

We have
\begin{equation}\label{exprW}
\left\{
\begin{aligned}
& W_A = 2 \kA \frac{m^2}{\rho^4} z \cos i,
\\
& W_B = \kB \frac{m \nu_0}{2\rho}
\left(
- \nu_0 z \cos i + \dot v z - \dot r \cos u \sin i
\right),
\\
& W_C = \kC \frac{m \nu_0}{\rho}
\left(
\frac{3}{2}  \nu_0 z \cos i - \dot v z + \dot r \cos u \sin i
\right),
\\
& W_{\widehat N} = k_N \frac{m \nu_0^2}{\rho}
\frac{3}{2}  z \cos i.
\end{aligned}
\right.
\end{equation}
Rearranging terms it follows
\begin{equation*}
\begin{aligned}
W = & \frac{m \nu_0^2}{\rho}
\left(
2 \kA - \frac{\kB}{2} + \frac{3 \kC}{2} + \frac{3 k_N}{2}
\right) z \cos i
\\
&
+ \frac{m \nu_0 \dot v}{\rho} \left(
\frac{\kB}{2} - \kC
\right) z
+ \frac{m \nu_0}{\rho} \left(
- \frac{\kB}{2} + \kC
\right) \dot r \cos u \sin i.
\end{aligned}
\end{equation*}
When
$\HH = -2$, $\FF = 2$
and $\II =0$ (i.e., $\beta = \gamma=1$)
and $t_1=t_2=t_3=0$,
 we find
$$
W_{{\rm GR}} = \frac{3m\nu_0}{\rho} z \dot v - \frac{3 m\nu_0}{\rho}
\dot r \cos u \sin i.
$$
When the satellite is the Moon, we have $\dot r \sin i
= O(e i)$ which is negligible. In this case
$W_{{\rm GR}}$ corresponds to the formula found by
de Sitter in
\cite[(95)]{dS:16-2}.

Now we compute the Gaussian component $S$ of the perturbation.
Analogously we decompose
$$
S = S_A + S_B + S_C + S_{\widehat N}.
$$
We have
\begin{equation}\label{exprS}
\left\{
\begin{aligned}
& S_A = - \kA \frac{m \nu_0^2}{r \rho} (r^2 - 2 z^2),
\\
&
S_B = \kB \frac{m \nu_0}{2r \rho}
\left[
r^2 \left(
\nu_0 - \dot v  \cos i\right) - \nu_0 z^2
\right],
\\
&
S_C = \kC \frac{m \nu_0}{r \rho}
\left[
r^2 \left(
-\frac{\nu_0}{2} + \dot v  \cos i\right)
+ \frac{3}{2} \nu_0 z^2
\right],
\\
& S_{\widehat N} = k_N \frac{m \nu_0^2}{2 r\rho}
\left(
-r^2 + 3 z^2
\right).
\end{aligned}
\right.
\end{equation}
Rearranging terms it follows
$$
\begin{aligned}
S= & \frac{m\nu_0}{r\rho}
\Bigg[
\left(
- \kA + \frac{\kB}{2} - \frac{\kC}{2} - \frac{k_N}{2}
\right) \nu_0 r^2
+
\left(
-
\frac{\kB}{2} + \kC
\right) \dot v r^2 \cos i
\\
& +
\left(
2 \kA - \frac{\kB}{2} + \frac{3}{2} \kC
+ \frac{3}{2} k_N
\right)
\nu_0 z^2
\Bigg].
\end{aligned}
$$
When
$\HH = -2$, $\FF = 2$
and $\II =0$ and $t_1=t_2=t_3=0$,
 we find
$$
S_{{\rm GR}} =
- 3 \frac{m\nu_0 \dot v}{\rho} r \cos i.
$$
When the satellite is the Moon if we approximate
 $\cos i \simeq 1$ then $S_{{\rm GR}}$ corresponds to
the formula found in
\cite[(95)]{dS:16-2}.

Similarly we decompose
$$
T = T_A + T_B + T_C + T_{\widehat N}.
$$
We have
\begin{equation}\label{exprT}
\left\{
\begin{aligned}
& T_A = 2 \kA \frac{m^2}{\rho^4} z \cos u \sin i,
\\
&
T_B
= \kB \frac{m \nu_0}{2\rho}
\left(
-\nu_0 z \cos u \sin i + \dot r \cos i
\right),
\\
&
T_C
= \kC \frac{m \nu_0}{\rho}
\left(
\frac{3}{2} \nu_0 z \cos u \sin i - \dot r \cos i
\right),
\\
& T_{\widehat N} = \frac{3}{2} k_N \frac{m \nu_0^2}{\rho}
z \cos u \sin i.
\end{aligned}
\right.
\end{equation}
Rearranging terms it follows
$$
\begin{aligned}
T=
&
\frac{m\nu_0}{\rho}
\Bigg[
\left(
2 \kA - \frac{\kB}{2} + \frac{3}{2} \kC + \frac{3}{2} k_N
\right) \nu_0 z \cos u \sin i
+
\left(
\frac{\kB}{2} - \kC
\right)
\dot r \cos i
\Bigg].
\end{aligned}
$$
When
$\HH = -2$, $\FF = 2$
and $\II =0$
and $t_1=t_2=t_3=0$,
 we find
$$
T_{{\rm GR}} =
 3 \frac{m\nu_0}{\rho} \dot r \cos i.
$$
When the satellite is the Moon if we approximate
 $\cos i \simeq 1$ then $T_{{\rm GR}}$ corresponds to
the formula found in
\cite[(95)]{dS:16-2}.

As for the computation of the precession of pericenter in Section
\ref{sec:comput},
we use the formulae
\begin{equation}\label{ra1e}
 r=
\frac{a(1-e^2)}{1+e\cos v},\qquad \qquad \dot r=\frac{ae(1-e^2)\dot{v}
\sin v }{(1+e\cos v)^2}.
\end{equation}
We also recall the following planetary equations of Lagrange in
the Gauss form
\cite[Ch. 6, Sec. 6]{GeWe:71}:
%
\begin{equation}\label{planetary}
\begin{aligned}
\frac{d\Omega}{dt}
=& \frac{1}{na^2 (1-e^2)^{1/2} \sin i} ~ W r \sin u,
\\
\\
\displaystyle
\frac{d\omt}{dt}= &
\frac{(1-e^2)^{1/2}}{nae}
\left[
-S\cos v +T\left(1+{r\over
a(1-e^2)}
\right)\sin v
\right]+
2\sin^2 \frac{i}{2} ~{d\Omega\over dt},
\end{aligned}
\end{equation}
where $n = 2\pi/U$, $U$ the period of revolution of the satellite
around Earth.

We make the computations up to the
 first order with respect to the eccentricity $e$, and we use
the following formula for the true anomaly
\cite[formula (2.6.7)]{GeWe:71}
%
\begin{equation}\label{vditi}
v(t) = n (t - \tau) + 2 e \sin \left[
n (t-\tau)
\right] + O(e),
\end{equation}
where $\tau$ is the satellite time of perigee passage.


\section{Precession of orbital elements in the presence
of torsion} \label{sec:corredue}
Using the expressions of $S$, $T$ and $W$ computed
in the previous section and
integrating \eqref{planetary} we find secular terms in the
expressions of
$\inttempo \Omega$ and $\inttempo \widetilde \omega$.
According to perturbation theory, we regard
the orbital elements as approximately constant
in the computation of such integrals. Since $u = v + \widetilde \omega-
\Omega$,
we can make use of the approximation $\dot u \simeq \dot v$. Moreover
we use the formula $na^2 (1-e^2)^{1/2} = r^2 \dot v$, and
\eqref{vditi}.

Let us first consider the computations for the node $\Omega$.
Substituting the decomposition
\eqref{exprW} of $W$  into the expression of $\frac{d\Omega}{dt}$
in \eqref{planetary}, one has to compute the following three types
of integrals:
$$
\mathrm{I} = \int \frac{z r \sin u}{n a^2 (1-e^2)^{1/2}} ~dt, \qquad
\mathrm{II} = \int \frac{z \dot v r \sin u}{n a^2 (1-e^2)^{1/2}} ~dt, \qquad
\mathrm{III} = \int \frac{r \dot r \sin u \cos u}{n a^2 (1-e^2)^{1/2}} ~dt.
$$
The integrals $\mathrm{I}$ and $\mathrm{II}$
yield periodic terms plus the following secular
contributions:
$$
\mathrm{I}_{\rm sec} = \frac{\sin i}{n} \left( t - \frac{v}{2n}\right),
\qquad
\mathrm{II}_{\rm sec} = \sin i \frac{t}{2},
$$
where $t$ is time.
The integral $\mathrm{III}$ yields only periodic terms.

In conclusion, it turns out that
the secular contributions to the variation of the
node $\Omega$ are:
\begin{equation}\label{secula}
(\inttempo \Omega)_{\rm sec} = \frac{m \nu_0^2}{\rho}
\left(
2 \kA - \frac{\kB}{2} + \frac{3 \kC}{2} +
\frac{3 k_N}{2}
\right)
\left(t - \frac{v}{2n}\right) \frac{\cos i}{n}
+
\frac{m \nu_0}{\rho} \left(
\frac{\kB}{2} - \kC
\right) \frac{t}{2}.
\end{equation}
Since $v=nt + {\rm periodic~terms~in}~v$,
inserting \eqref{eq:KaKbKcKd} and \eqref{eq:kN}
into \eqref{secula} we obtain
\begin{equation}
\begin{aligned}
(\inttempo \Omega)_{\rm sec} = &
\frac{1}{2}
\frac{m \nu_0}{\rho}
\Bigg\{
- \frac{\HH}{2} + \FF + \frac{\torparam_1}{2} + \frac{3\torparam_2}{2}
\\
& - \frac{\nu_0}{2n} \cos i
\left[
\frac{\HH}{2} (\HH + \FF) - \II - \torparam_1 \FF + \torparam_1 + \torparam_2 + 2 \torparam_3
\right]
\Bigg\} ~t.
\end{aligned}
\end{equation}
Using the Newtonian limit \eqref{limiteniutoniano} and setting
\begin{equation}\label{cuno}
\costante_1  \equiv 1-  \frac{\HH}{2}  + 2 \FF + 3 \torparam_2,
\qquad
\costante_2  \equiv 1 + \frac{\HH}{2} + \frac{\HH^2}{2}
- \FF - \II + \torparam_2 + 2 \torparam_3,
\end{equation}
we obtain
\begin{equation}\label{marrazzzzo}
(\inttempo \Omega)_{\rm sec} =
\frac{1}{4}
\frac{m \nu_0}{\rho}
\left(
\costante_1 - \costante_2 \frac{\nu_0}{n} \cos i
\right)
~t.
\end{equation}
\noindent When torsion is zero, that is $\torparam_1= \torparam_2=
\torparam_3=0$, if we let
$\FF=2\gamma$, $\II=2(\beta-\gamma)$ and $\HH=-2$
(PPN formalism) we obtain
\begin{equation}\label{finke}
(\inttempo \Omega)_{\rm sec} =
\frac{1}{2}
\frac{m\nu_0}{\rho}
\left( 1 +
2 \gamma\right) t
-
\frac{1}{2}
\frac{m\nu_0}{\rho}
\left(1-\beta\right)
\frac{\nu_0}{n} \cos i~
 t.
\end{equation}
The first term on the right hand side of \eqref{finke}
determines the usual geodetic precession effect. The second
term is consistent with the computations in
\cite[formula (48A)]{FiKr:76},
when the satellite is the Moon, so that we can
approximate $\cos i \simeq 1$.

Letting $\gamma= \beta =1$ we find
the usual formula of geodetic
precession in General Relativity found by de Sitter
(see
\cite[formula (97)]{dS:16-2}),
$$
(\inttempo \Omega)_{\rm sec}^{{\rm GR}}
=
\frac{3 m\nu_0}{2\rho} t.
$$
We recall that when the satellite is the Moon, this precession amounts  to $1.92~{\rm arcsec}/{\rm century}$.

Now we consider the computations for the perigee.
The contribution of the gaussian component $S$ to
the variation $\delta \widetilde \omega$ of the perigee
is given by the integral
\begin{equation}\label{contrStildew}
- \frac{(1-e^2)^{1/2}}{nae}
\int
S \cos v ~dt.
\end{equation}

Substituting the decomposition
\eqref{exprS} of $S$  into the above integral,
one has to compute the following three types
of integrals:
$$
\mathrm{I}^S = \int r \cos v~dt,
\qquad
\mathrm{II}^S = \int r \dot v \cos v~dt,
\qquad
\mathrm{III}^S = \int \frac{z^2}{r} \cos v~dt.
$$
We evaluate such integrals for small values of the eccentricity
$e$ and, taking into account that $e$ appears at the denominator
of \eqref{contrStildew}, we expand the integral up to the second
order in $e$. This is accomplished by expanding $r$ in \eqref{ra1e}
with respect to $e$, and inserting \eqref{vditi} in the resulting expression.
Then each of the integrands of
$\mathrm{I}^S$, $\mathrm{II}^S$, $\mathrm{III}^S$ turns out to be a sum of
products of simple trigonometric functions, from which secular
terms and periodic terms can be separated.
Such integrals then yield the following secular
contributions:
$$
\mathrm{I}^S_{\rm sec} =
- \frac{3}{2} a e
(1-e^2) t,
\qquad
\mathrm{II}^S_{\rm sec} =
- \frac{1}{2} a e
(1-e^2) v,
$$
and
\begin{equation}\label{IIISsec}
\mathrm{III}^S_{\rm sec} =
- \frac{3}{8} a e
(1-e^2) \sin^2 i
\left[
3 \sin^2 \left(\widetilde \omega- \Omega\right)
+
\cos^2 \left(\widetilde \omega- \Omega\right)
\right] t.
\end{equation}
In order to evaluate the integral $\mathrm{III}^S$,
since $z = r \sin u \sin i$,
we have made use
of the relation $u = v + \widetilde \omega
- \Omega$ to express the trigonometric functions of $u$.

The contribution of the gaussian component $T$ to
the variation $\delta \widetilde \omega$ of the perigee
is given by the integral
\begin{equation*}
 \frac{(1-e^2)^{1/2}}{nae}
\int
T \left(
1 + \frac{r}{a(1-e^2)}
\right)
\sin v ~dt.
\end{equation*}
Substituting the decomposition
\eqref{exprT} of $T$  into the above integral,
one has to compute the following two types
of integrals:
$$
\mathrm{I}^T =
\int
\left(
1 + \frac{r}{a(1-e^2)}
\right)
z \cos u \sin v
~dt,
\qquad
\mathrm{II}^T =
\int
\left(
1 + \frac{r}{a(1-e^2)}
\right)
\dot r \sin v
~dt.
$$
As in the previous case, we evaluate such integrals for small values
of the eccentricity $e$. Moreover, we will use the expression in
\eqref{ra1e} for $\dot r$ and the relation $u = v + \widetilde \omega - \Omega$.

Then each of the integrands of
$\mathrm{I}^T$, $\mathrm{II}^T$ turns out to be a sum of
products of simple trigonometric functions,
and we can extract
the following secular
contributions:
\begin{equation*}
\begin{aligned}
\mathrm{I}^T_{\rm sec} = &
\frac{7}{8} a e
(1-e^2) \sin i
\left[
\sin^2 \left(\widetilde\omega -\Omega
\right) - \cos^2
\left(\widetilde \omega - \Omega\right)
\right]
t,
\\
\mathrm{II}^T_{\rm sec} = &
a e
(1-e^2) v.
\end{aligned}
\end{equation*}
In conclusion,
the secular contribution to the variation of the
perigee $\widetilde \omega$ is:
\begin{equation}\label{seculaomega}
\begin{aligned}
(\inttempo \widetilde \omega)_{\rm sec} =
&
\frac{1}{2}
\frac{m \nu_0}{\rho}
\left(
\frac{\kB}{2} - \kC
\right) t
+ \frac{3}{2}
\frac{m \nu_0^2}{n \rho}
\left(
- \kA + \frac{\kB}{2} - \frac{\kC}{2} - \frac{k_N}{2}
\right) t
\\
& + \frac{1}{2} \frac{m \nu_0^2}{n \rho}
\left[
5 \sin^2 i \sin^2 (\widetilde \omega - \Omega)
- (1-\cos i)
\right] \left(
2 \kA - \frac{\kB}{2} + \frac{3}{2} \kC + \frac{3}{2} k_N
\right) t.
\end{aligned}
\end{equation}
Inserting \eqref{eq:KaKbKcKd} and \eqref{eq:kN}
into \eqref{seculaomega} we obtain
$$
\begin{aligned}
(\inttempo \widetilde \omega)_{\rm sec} = &
\frac{1}{2}
\frac{m \nu_0}{\rho}
\left(
- \frac{\HH}{2} + \FF + \frac{\torparam_1}{2} + \frac{3
\torparam_2}{2}
\right) t
+ \frac{3}{4}
\frac{m \nu_0^2}{n \rho}
\left(
- \HH - \FF - \II
+ \torparam_1 + \torparam_2 + 2 \torparam_3 + \torparam_1 \HH
\right) t
\\
& - \frac{1}{4} \frac{m \nu_0^2}{n \rho}
\left[
5 \sin^2 i \sin^2 (\widetilde \omega - \Omega)
- (1-\cos i)
\right] \left(
\frac{\HH}{2}(\HH + \FF)
- \II - \torparam_1 \FF + \torparam_1 + \torparam_2 + 2
\torparam_3
\right) t.
\end{aligned}
$$
Using \eqref{limiteniutoniano} and \eqref{cuno}
we get
\begin{equation}\label{marrazzopiero}
(\inttempo \widetilde \omega)_{\rm sec} =
\frac{1}{4} \frac{m \nu_0}{\rho}
\left\{
\costante_1  + \costante_2 \frac{\nu_0}{n}
\left[
4 - \cos i - 5 \sin^2 i \sin^2 (\widetilde \omega - \Omega)
\right]
\right\} ~t.
\end{equation}

When torsion is zero, in the PPN formalism we obtain
\begin{equation}\label{finkebis}
\begin{aligned}
(\inttempo \widetilde \omega)_{\rm sec} = &
\frac{1}{2}
\frac{m \nu_0}{\rho}
\left(
1+2 \gamma\right)~t
 +  \frac{3}{2}
\frac{m \nu_0}{\rho}
(1-\beta) \frac{\nu_0}{n}
~ t
\\
& - \frac{1}{2} \frac{m \nu_0^2}{n \rho}
\left[
5 \sin^2 i \sin^2 (\widetilde \omega - \Omega)
- (1-\cos i)
\right] (1-\beta) t.
\end{aligned}
\end{equation}
The first term on the right hand side of \eqref{finkebis}
determines the usual geodetic precession effect. The second
term is consistent with the computations in
\cite[formula (48A)]{FiKr:76},
when the satellite is the Moon, so that we can
approximate $\cos i \simeq 1$ (and the third term becomes negligible).

Letting $\gamma= \beta =1$ we find
$C_1=6$, $C_2=0$, hence
 the usual formula of geodetic
precession in General Relativity found by de Sitter
(see
\cite[formula (97)]{dS:16-2}),
$$
(\inttempo \widetilde \omega)_{\rm sec}^{{\rm GR}}
=
\frac{3 m\nu_0}{2\rho} t.
$$
Both in $(\delta \Omega)_{\rm sec}$ and in $(\delta \widetilde \omega)_{
\rm sec}$ there appears the term
$$
\frac{1}{4}
\frac{m \nu_0}{\rho}
\left(1-  \frac{\HH}{2}  + 2 \FF + 3 \torparam_2\right) t,
$$
which is independent of $n$ and thus of the details of the satellite's
motion. This term can therefore be interpreted as the geodetic
precession effect when torsion is present.

\section{Torsion biases}\label{sec:torbia}
In this section, similarly to
\cite{MaTeGuCa:07},
we define
multiplicative torsion biases relative to the GR predictions.
These torsion biases will be used to put constraints on torsion
parameters from solar system experiments.

For the case of  precession of the satellite orbital elements
we define
\begin{equation}\label{miarrazzo}
\begin{aligned}
b_{\,\Omega} \equiv &
\frac{(\inttempo \Omega)_{{\rm sec}}}{(\inttempo \Omega)_{{\rm sec}}^{{\rm GR}}},
\qquad\qquad
b_{\,\widetilde \omega} \equiv &
\frac{(\inttempo \widetilde
\omega)_{{\rm sec}}}{(\inttempo \widetilde \omega)_{{\rm sec}}^{{\rm GR}}}.
\end{aligned}
\end{equation}
{}From \eqref{marrazzzzo} and \eqref{marrazzopiero} it follows
\begin{equation}\label{marrazzochefaistasera?}
\begin{aligned}
b_{\,\Omega} = & \frac{1}{6}
\left(C_1 - C_2 \frac{\nu_0}{n} \cos i\right),
\\
\\
b_{\,\widetilde \omega} = & \frac{1}{6}
\left\{C_1 + C_2 \frac{\nu_0}{n} \left[
4-
\cos i
- 5 \sin^2 i \sin^2 \left(
\widetilde \omega - \Omega
\right)
\right]
\right\},
\end{aligned}
\end{equation}
where the constants $C_1$ and $C_2$ are defined in \eqref{cuno}.

For the purpose of comparison of solar system experiments
with the predictions from the present torsion theory, we will assume
that all metric parameters take the same form as in PPN formalism,
according to \eqref{paramPPN}. We have
\begin{equation}
\begin{cases}
C_1 = 2 + 4 \gamma + 3 \torparam_2,
\\
C_2 = 2 - 2\beta + \torparam_2 + 2 \torparam_3,
\end{cases}
\end{equation}
and from (\ref{marrazzochefaistasera?}) we obtain
\begin{equation}\label{marrazzovainconvento}
\begin{cases}
b_{\,\Omega} =
\displaystyle
\frac{1}{3} (1 + 2\gamma)
+\frac{\torparam_2}{2} -
\frac{\nu_0}{3n} \cos i \left(
1 - \beta + \frac{\torparam_2}{2} + \torparam_3
\right),
\\
\\
b_{\, \widetilde \omega} =
\displaystyle
\frac{1}{3} (1 + 2\gamma)
+\frac{\torparam_2}{2} +
\frac{\nu_0}{3n} \left[
4 - \cos i - 5 \sin^2 i \sin^2 \left(
\widetilde \omega - \Omega\right)
\right] \left(
1 - \beta + \frac{\torparam_2}{2} + \torparam_3
\right).
\end{cases}
\end{equation}
Note that the torsion correction to the geodetic precession in
\eqref{marrazzovainconvento}
(namely the term
$\frac{1}{3}(1+2\gamma) + \frac{t_2}{2}$
) differs from the corresponding one found for gyroscopes in
\cite[formula (47)]{MaTeGuCa:07}
by a numerical factor of order of unity.

When the satellite is the Moon, we have $\cos i = 1 + O(i^2)$,
$\sin^2 i = O(i^2)$, and we may approximate
\begin{equation}\label{marrazzovasullaluna}
\begin{cases}
b_{\,\Omega} =
\displaystyle
\frac{1}{3} (1 + 2\gamma)
+\frac{\torparam_2}{2} -
\frac{\nu_0}{3n} \left(
1 - \beta + \frac{\torparam_2}{2} + \torparam_3
\right),
\\
\\
b_{\, \widetilde \omega} =
\displaystyle
\frac{1}{3} (1 + 2\gamma)
+\frac{t_2}{2} +
\frac{\nu_0}{n} \left(
1 - \beta + \frac{t_2}{2} + t_3
\right).
\end{cases}
\end{equation}
Similarly, considering now $\delta \widetilde \omega$
as the precession of pericenter computed in Section
\ref{sec:comput}, we have
\begin{equation}\label{mrrzz}
B_{\widetilde \omega}
\equiv \frac{ (\delta \widetilde \omega)_{\rm sec}}{(\delta \widetilde \omega)^{\rm GR}_{\rm sec}}
=
\frac{1}{3}
\left(
1 + \frac{\HH^2}{4} + \frac{\FF}{2} - \frac{\II}{2} + 2 t_2
+ t_3
\right).
\end{equation}
In the case of a PPN metric we find
\begin{equation}\label{marrazzoelautoblu}
B_{\widetilde \omega} = \frac{1}{3} \left(
2 + 2\gamma - \beta + 2 t_2 + t_3\right).
\end{equation}
%

\section{Constraining torsion with the Moon and Mercury}\label{sec:limits}

Here we compare the predicted torsion biases to experimental
measurements in order to set limits on the torsion parameters.

Recent limits on various components of the torsion tensor, obtained
in a different torsion model based on the fact that
background torsion may violate effective local Lorentz invariance,
have been
obtained in \cite{KoRuTa}. See also \cite{HeAd}, where
constraints on possible new spin-coupled interactions using
a torsion pendulum are described.

\subsection{Moon: geodetic precession}

The GR test of the geodetic precession, evaluated with LLR data and
expressed as a relative deviation from the value expected in GR, is
$K_{gp} = -0.0019 \pm 0.0064$
\cite{Williams2004}.
In our torsion theory this
is to be compared with the first two terms
on the right hand side
of
equation \eqref{marrazzovainconvento}:
\begin{equation}
|~\widehat b_{\, \widetilde \omega} - 1| =
\displaystyle
|\frac{1}{3} (1 + 2\gamma) + \frac{t_2}{2} - 1| =
|\frac{2}{3} (\gamma - 1) + \frac{t_2}{2}| < 0.0064,
\end{equation}
where, taking into account the last sentence
at the end of Section \ref{sec:corredue}, we define
$$
\widehat b_{\widetilde \omega} \equiv \frac{1}{3} (1 + 2\gamma)
+ \frac{t_2}{2}.
$$
Using the Cassini measurement $\gamma - 1 = (2.1 \pm 2.3) \times 10^{-5}$,
we can neglect
this term compared to the experimental uncertainty on $K_{gp}$ and get
the following constraint on $t_2$:
\begin{equation}
\label{MoonLimitOn-t2}
2~ |~ \widehat b_{\, \widetilde \omega} - 1| =
\displaystyle
|t_2| < 0.0128.
\end{equation}
The meaning of the constraint on the torsion parameter $t_2$ is the following.
Using the value $(\inttempo \widetilde \omega)_{{\rm sec}}^{{\rm GR}}=1.92$ arcsec/century
for the geodetic precession of the Moon's orbit in GR, the geodetic precession in the presence of
torsion is
$$
(\inttempo \widetilde\omega)_{{\rm sec}}=
\left[1+\frac{2}{3} (\gamma - 1) + \frac{t_2}{2}\right]1.92\quad\mbox{arcsec/century}.
$$
If the parameter $t_2$ had a value larger than 0.0128, this would imply the precession
of the Moon's perigee would be (neglecting the contribution of $\gamma - 1$)
$$
(\inttempo \widetilde\omega)_{{\rm sec}}>
(1+0.0064)1.92\quad \mbox{arcsec/century},
$$
which would be inconsistent with LLR data. If $t_2<-0.0128$ we would have an analogous inconsistency. To give the reader a further feeling of the effect of a nonzero value of $t_2$ in terms of orbit displacement, we provide the following, extreme example. The 1.92 arcsec/century precession amounts to a perigee displacement of about 3 meters per lunar orbit period (about 27 days). A value $t_2 = 1$ would imply (neglecting the contribution of $\gamma - 1$) a geodetic precession in the presence of torsion of about 4.5 meters/lunar orbit period.

The relentless accumulation of LLR data, the mm range accuracy of the APOLLO
station, the start or restart of LLR operation of additional ILRS\footnote{International
Laser Ranging Service; see http://ilrs.gsfc.nasa.gov/.} stations
(like MLRO,
the Matera Laser Ranging Observatory in the south of Italy) will provide
continuous
further improvements of the $K_{gp}$ test and therefore, of this limit on $t_2$.
Future improvements are possible also with current data and
stations, by further developing and refining the current orbital software
packages.

\subsection{Mercury: perihelion advance}
The perihelion advance of Mercury has been measured with
planetary radar ranging by Shapiro et al. in 1989
\cite{shapiro-mercury}.
They found it to be consistent with its GR value with a relative standard
error of $10^{-3}$. In the PPN framework, this can be used to ``infer
that $\beta = 1$ to within a standard error of $\sigma(\beta) = 0.003$''
(quoted from
\cite{shapiro-mercury}).

According to our torsion model, in the case of the PPN metric with
the torsion bias given by
\eqref{marrazzoelautoblu}, we get for Mercury the one standard deviation limit:
\begin{equation}
\begin{aligned}
|B_{\widetilde \omega} -1| = \frac{1}{3} |2\gamma - \beta -1 + 2 t_2 + t_3|
< 0.001,
\\
\\
|2\gamma - \beta -1 + 2 t_2 + t_3|<0.003,
\\
\\
|(2\gamma - 2) + (1 - \beta) + 2 t_2 + t_3|<0.003.
\end{aligned}
\end{equation}
Using the Cassini measurement $\gamma - 1 = (2.1 \pm 2.3)\times 10^{-5}$,
we can neglect
this term compared to the experimental uncertainty on the perihelion advance
and get the constraint:
\begin{equation}
\label{MercuryLimitOn-beta-t2-t3}
|1 - \beta + 2 t_2 + t_3|<0.003.
\end{equation}

\begin{figure}[!ht]
\begin{center}
  \includegraphics[width=0.9\textwidth]{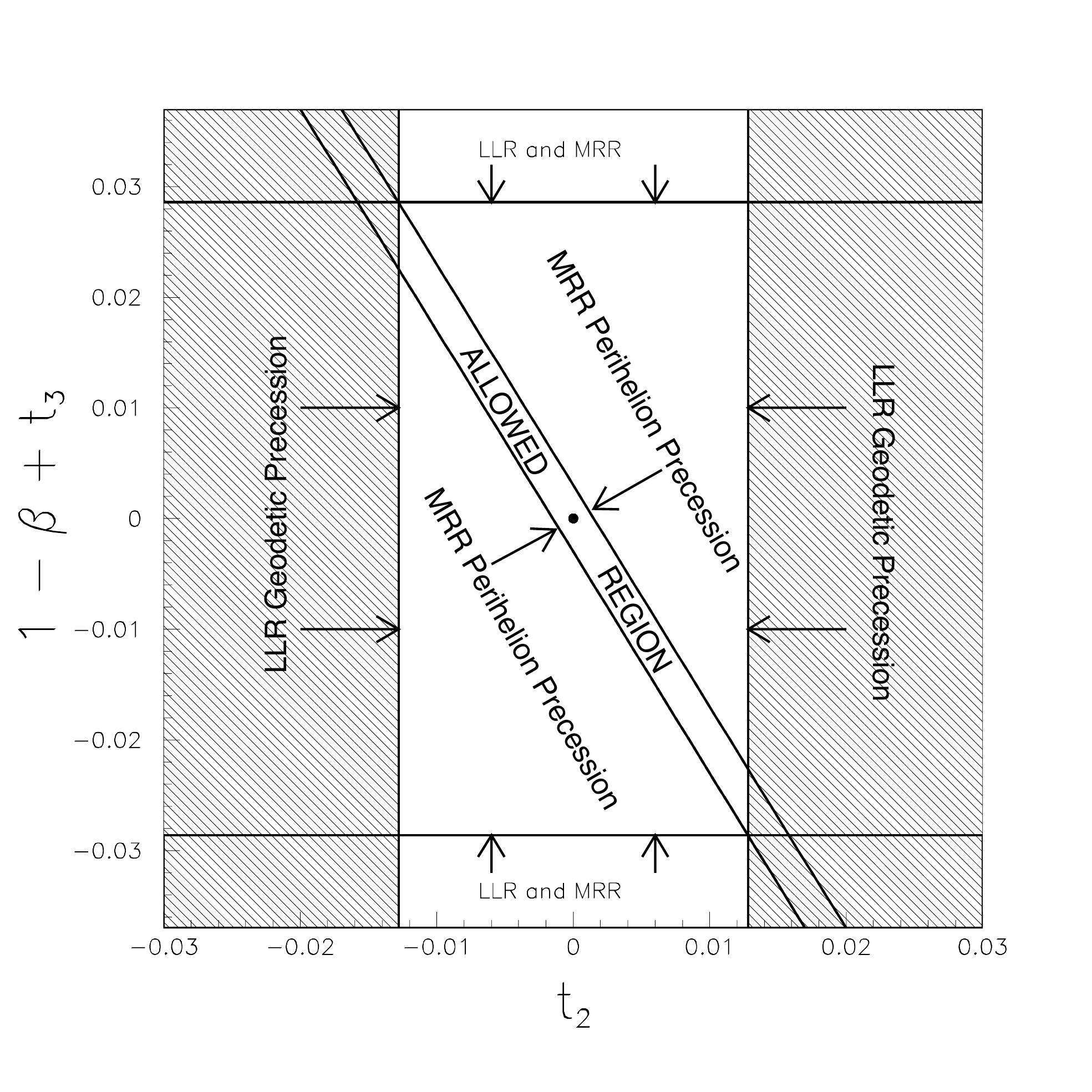}
\caption{{\small Constraints on  $t_2$ and the linear combination  $(1-\beta)+t_3$ from LLR and
MRR, indicated by the arrows. For example, the hatched area is the region
excluded by LLR only. General Relativity corresponds to $\beta -1= t_2 = t_3 = 0$ (black dot).}}
\label{plot}
\end{center}
\end{figure}

The limits \eqref{MoonLimitOn-t2} and \eqref{MercuryLimitOn-beta-t2-t3} on the
values of $t_2$ and $(1-\beta) + t_3$ are represented graphically in fig. \ref{plot}.

Combining the LLR and MRR constraints one gets on $t_3$ the following limit:
\begin{equation}
\label{CombinedLimitOn-t3_old}
|1-\beta + t_3| < 0.0286.
\end{equation}
If we assume the Nordtvedt effect
\cite{nordtvedt-effect}, and that the Nordtvedt parameter $\eta_N^{} = 4\beta - \gamma - 3$,
then the measured value \cite{Williams2004}  is $\beta = 1 +  (1.2 \pm 1.1)\times 10^{-4}$,
which makes $\beta-1$ negligible compared to the experimental uncertainty
on the perihelion advance. The constraint then becomes:
\begin{equation}
\label{MercuryLimitOn-t2-t3}
|2t_2 + t_3|<0.003.
\end{equation}
In this latter case, combining the LLR and MRR constraints one gets on $t_3$ the following limit:
\begin{equation}
\label{CombinedLimitOn-t3}
|t_3| < 0.0286.
\end{equation}
The meaning of the constraints on the torsion parameter $t_3$ is the following.
Using the value $(\inttempo \widetilde \omega)_{{\rm sec}}^{{\rm GR}}=42.98$ arcsec/century
for the precession of Mercury's perihelion in GR, the precession of the perihelion in the presence of
torsion is (neglecting the contribution of $\gamma - 1$):
$$
(\inttempo \widetilde\omega)_{{\rm sec}}=
\left[1+\frac{1}{3} \left(1 - \beta + 2 t_2 + t_3\right)\right]42.98\quad\mbox{arcsec/century}.
$$
If the linear combination
$(1 - \beta + 2 t_2 + t_3)$ had a value larger than 0.003, this would imply the precession
of Mercury's perihelion would be
\begin{equation}
\label{inconsistency}
(\inttempo \widetilde\omega)_{{\rm sec}}>
(1+0.001)42.98\quad \mbox{arcsec/century},
\end{equation}
which would be inconsistent with MRR data. If the parameter $t_2$ takes
the lowest value consistent with LLR data, i.e., if $t_2=-0.0128$, then
a value of the parameter $t_3$ larger than $(0.0286 + \beta - 1)$ would imply the
inconsistency (\ref{inconsistency}) with MRR data. If $\beta - 1$ is neglected, then simply
$t_3>0.0286$ would be inconsistent with the data.
Eventually, if $(1 - \beta + 2 t_2 + t_3) < -0.003$ we would have analogous inconsistencies.

We stress that the perihelion advance measurement used here is based on data
taken between 1966 and 1990. As pointed out by Will (\cite{Will},  page 38)
``analysis of data taken since 1990 could improve the accuracy''. Therefore,
the above contraints on spacetime torsion can be improved already now, with
existing data, while waiting that Mercury is reached by new spacecrafts, like
in particular ESA's BepiColombo.

\section{Equations of geodesics}\label{marrazzosifaprete}
In this section we consider the particular case of geodesic trajectories.
The system of equations of geodesics trajectories reads as
$$
\frac{d^2 x^{\primoindice}}{d\tau^2} +
\chr{\primoindice}{\secondoindice\terzoindice}
\frac{d
x^{\secondoindice}}{d\tau}\frac{d x^{\terzoindice}}{d\tau}=0.
$$
The resulting system of equations of motion is given by \eqref{sistODEsferiche}
with $\torparam_1=\torparam_2=\torparam_3=0$.
Such a system to lowest order becomes
\begin{equation*}
\frac{d \vec v}{dt} = \frac{\HH}{2} \frac{m}{r^2} \hat e_r.
\end{equation*}
Imposing the Newtonian limit (see also
\cite[formula (25)]{MaTeGuCa:07})
it follows that
$$
\HH = -2.
$$
Hence all the precession formulae are the same as in
the PPN formalism given in \eqref{formulaPPN}, \eqref{finke}
and \eqref{marrazzopiero}.
Therefore, if a satellite's orbit is assumed to be a geodesic, then
 the measurements of satellite experiments cannot be used
to constrain the torsion parameters.

\section{Discussion of the results and future prospects}\label{sec:discu}
This work is an investigation of the effects of spacetime
torsion on the orbits of satellites and planets,
based on a model with several parameters evolved
from the one by MTGC. Due to the presence of a set  of parameters,
it must be tested with a combination of experiments designed to measure
different physical effects and observables. In this case, no single
experiment provides a complete answer, but experiments with the
best accuracy and the broadest parameter sensitivity may find the
first reliable hint of torsion. The notable example is the geodetic precession,
which can be measured using three very different instrumental techniques:
LLR, GPB's gyroscopes and future BepiColombo's radio science and
accelerometer payloads. This makes
constraining torsion with the geodetic precession robust against the
effect of experimental systematic errors.

For completeness, we quote here   that  the
constraints on torsion provided by the
 measurement of the geodetic
precession turn out to be useful also in constraining spacetime torsion with
the frame dragging experiments on  LAGEOS satellites.
We remark
that the torsion corrections to
the Lense-Thirring effect for LAGEOS and GPB contain different sets of torsion
parameters. We refer to the companion paper \cite{LT}
for the details.

\subsection{Future prospects}\label{sub:future}
Before the end of the decade, robotic missions on the lunar surface could
deploy new scientific payloads which include laser retroreflectors
and thus extend the LLR reach for new physics in three ways: (i) using a significantly improved
2$^{nd}$ generation retroreflector design; (ii) increasing by a factor about 2 the
geometric lever arm of LLR with missions to the lunar poles or limb; (iii) combining
LLR payloads with transponders (at least two) for same-beam microwave interferometry (SBI) capable
of additional accurate measurements of lunar rotations and librations
\cite{SBI}, \cite{Bender}
(during the lifetime of the transponders).

In particular, the single, large, fused-silica retroreflector design developed
by the University of Maryland and INFN-LNF
\cite{IAC}
will improve over the performance of current Apollo arrays
by a factor 100 or more,
 thus removing the dominant contribution
to the LLR error budget.
Such a contribution is of the order of 2 cm.
It is
due to the multi-retroreflector structure of the arrays coupled to the
librations and rotations of the Moon with respect to the Earth.
The functionality of this specific new design, which inherits and is evolved
from the successful Apollo 11, 14 and 15 experience, is being validated by
thermal-vacuum-optical testing laboratory-simulated space conditions
at the INFN-LNF ``Satellite/lunar laser ranging Characterization Facility (SCF)''
\cite{ILRS08}, \cite{INFNPaper:10}, \cite{MAGIA}.

 Other instruments, like the seismometer and
the heat-flow probe, will provide important information to evaluate the LLR
systematic errors related to the environmental conditions of the
lunar surface and sub-surface layers of the lunar regolith.

After the end of this decade, results from the
BepiColombo Mercury orbiter, an ESA
Cornerstone mission equipped with a high-accuracy radio science and accelerometer
payloads to test GR, is expected to
improve the classical test of the perihelion advance
\cite{Milani}, \cite{Ashby}
(42.98~{\rm arcsec}/{\rm century}).
The latter measurement can be cross-checked by new MRR data taken
simultaneously with BepiColombo's ranging data (possibily, by the same ground stations).
In addition, we note that in the past a Mercury orbiter like
BepiColombo has been considered also for yet
another independent measurement of the
geodetic precession
\cite{Slava}, \cite{Slava1}
($20.50~{\rm arcsec}/{\rm century}$; to be compared to the $1.92 ~{\rm arcsec}/{\rm century}$
for the Moon). Mercury's special role in the search for new physics effects, and for spacetime
torsion in particular, is due to the relatively large value of its eccentricity
and to its short distance to the Sun.

In conclusion, using current LLR and MRR data we have set constraints on the
 dimensionless torsion parameter $t_2$ and the linear combination $(1-\beta)+t_3$ at $10^{-2}$ level.
In the future, LLR, MRR, GPB and, ultimately, BepiColombo together can
exclude non-zero values of $t_2$ and $t_3$ with accuracies significantly
below 1\%.

\section{Appendix}\label{sec:app}
In this appendix we compute the precession of pericenter using the method
in Appendix C of the online version
\cite{MaTeGuCa:07online}.

The $\lambda=t$ component of \eqref{eq:autopara} is
\begin{equation*}
\begin{aligned}
& \frac{d^2 t}{d \tau^2} +
\left( \Gamma^t_{~tr}
+ \Gamma^t_{~rt}
\right)
\frac{dr}{d\tau} \frac{dt}{d\tau} =0,
\end{aligned}
\end{equation*}
which yields
$$
\frac{d^2 t}{d \tau^2} +
\left[
\torparam_1 - \HH + \left(
\HH^2 - 2 \II + 2 \torparam_3\right)
\frac{m}{r}
\right]
\frac{m}{r^2} \frac{dr}{d\tau} \frac{dt}{d\tau} =0,
$$
from which one finds
\begin{equation*}
\frac{d}{d\tau}
\left[
\exp\left(\int \psi(r) ~dr\right)
\frac{dt}{d\tau}
\right]=0,
\end{equation*}
where
\begin{equation*}
\psi(r) \equiv
\left[
\torparam_1 - \HH + \left(
\HH^2 - 2 \II + 2 \torparam_3\right)
\frac{m}{r}
\right] \frac{m}{r^2}.
\end{equation*}
It follows the conservation law
\begin{equation}\label{eq:consene}
k \equiv
\exp\left\{
-\left[
\torparam_1 - \HH + \frac{1}{2}\left(
\HH^2 - 2 \II + 2 \torparam_3\right)
\frac{m}{r}
\right]\frac{m}{r}
\right\}
\frac{dt}{d\tau}
= {\rm const}.
\end{equation}
The $\lambda=\phi$ component of \eqref{eq:autopara} when $\theta = \pi/2$
is
\begin{equation*}
\frac{d^2 \phi}{d \tau^2} + \left(\Gamma^\phi_{~r\phi}+
\Gamma^\phi_{~\phi r}\right)
\frac{d r}{d\tau} \frac{d\phi}{d\tau} =0,
$$
which yields
$$
\frac{d^2 \phi}{d \tau^2} + \left(
\frac{2}{r}
- \torparam_2 \frac{m}{r^2}
\right)
\frac{d r}{d\tau} \frac{d\phi}{d\tau} =0,
\end{equation*}
where we can neglect the term with the factor $-2 t_4 \frac{m^2}{r^3}$.
It follows the conservation law
\begin{equation}\label{momento}
h \equiv r^2 \frac{d\phi}{d\tau} \exp\left(
\torparam_2 \frac{m}{r}
\right) = {\rm const}.
\end{equation}
Notice that
$k^2-1$ and $h^2$
are of the order $\em$.
Since the parameter $\tau$ is the proper time we have
\begin{equation*}
\frac{d s^2}{d\tau^2} = g_{\mu\nu} \frac{dx^\mu}{d\tau} \frac{dx^\nu}{d\tau} =
- 1,
\end{equation*}
from which, for a test body in the equatorial plane $\theta = \pi/2$
it follows
\begin{equation}\label{eq:metricains}
 - \left(
1+ \HH \frac{m}{r} + \II \frac{m^2}{r^2}
\right)
\left(\frac{dt}{d\tau}\right)^2
+
\left(
1+ \FF \frac{m}{r} \right)
\left(\frac{dr}{d\tau}\right)^2 +  r^2 \left(
\frac{d\phi}{d\tau}\right)^2= - 1.
\end{equation}
Observe that the term with the factor $\II$ is missing in
\cite[(C23)]{MaTeGuCa:07online}.

Using \eqref{eq:consene}, \eqref{momento} and the
identity $dr/d\tau = (dr/d\phi) (d\phi/d\tau)$, from \eqref{eq:metricains}
it follows
\begin{eqnarray}\label{curly}
\left(
\frac{dr}{d\phi}
\right)^2
=
\frac{r^4 \exp(2 \torparam_2 \frac{m}{r})}{h^2 \left(
1 + \FF \frac{m}{r}
\right)} &\Bigg\{
& -1 +
\frac{
k^2 \left(
1 + \HH \frac{m}{r} + \II \frac{m^2}{r^2}
\right)
}{
\exp\left\{
-2\left[
\torparam_1 - \HH + \frac{1}{2}\left(
\HH^2 - 2 \II + 2 \torparam_3\right)
\frac{m}{r}
\right]\frac{m}{r}
\right\}
}
\nonumber
\\
&&
- \frac{h^2}{r^2 \exp(2 \torparam_2 \frac{m}{r})}
\Bigg\}.
\end{eqnarray}
We now need to expand the right hand side of \eqref{curly} up to
the order $\em$, since
the left hand side is of order $O(1)$.
The
right hand side is divided by $h^2$, therefore we need to develop the
quantity in $\{\cdots\}$ up to the order $\em^2$.
Since
$k^2 - 1$ is of the order $\em$,
it is enough to
develop the exponential in front of $\{\cdots\}$
 up to the order $\em$.
Taking into account that $h^2$ is of order $\em$,
we have
\begin{eqnarray}\label{epmquadro}
\left(
\frac{dr}{d\phi}
\right)^2
=
\frac{r^4}{h^2}
\left[
1+ (2\torparam_2 - \FF)
\frac{m}{r}
\right]
&\Bigg\{
& -1 + k^2 + k^2 \left(2 \torparam_1 - \HH\right) \frac{m}{r}
\\
&& + k^2
\left[
2 \torparam_1 (\torparam_1- \HH) +
\HH^2 -  \II + 2 \torparam_3
\right]
\frac{m^2}{r^2}
- \frac{h^2}{r^2} + 2 \torparam_2 h^2 \frac{m}{r^3}
\Bigg\}.
\nonumber
\end{eqnarray}
{}From \eqref{epmquadro}, setting $u \equiv 1/r$, we obtain  the
differential equation of the orbit
\begin{equation}\label{essione}
\frac{d^2 u}{d\phi^2} + u =
\frac{m}{2h^2}
\left[
k^2 \left(
- \HH - \FF + 2 \torparam_1 + 2 \torparam_2
\right) + \FF - 2\torparam_2
\right]
+\frac{3}{2} \FF m u^2
+
\frac{k^2}{h^2} A m^2 u,
\end{equation}
where
\begin{equation}\label{eq:A}
A \equiv \HH \FF + \HH^2 - \II
- 2 \torparam_1 (\HH + \FF) + 2 \torparam_1^2 + 4
\torparam_1 \torparam_2 -
2 \torparam_2 \HH + 2 \torparam_3,
\end{equation}
where we have neglected the term containing $u^3$, which is
of the order $\em^2$.

We stress that the term
$\frac{k^2}{h^2} A m^2 u$,
neglected in the
computations of
\cite{MaTeGuCa:07},
is of order $\em$, as well as
the term
$\frac{3}{2} \FF m u^2$. Note also that $A=0$ in the case
of GR, and that $A = 2(2-\gamma - \beta)$ in the case of PPN.

Using \eqref{limiteniutoniano} and the above mentioned order
of magnitude of $k^2$ it follows
\begin{equation}\label{prec}
\frac{d^2 u}{d\phi^2} + \left(1 - \frac{A}{h^2}m^2
\right) u = \frac{m}{h^2} + \frac{3}{2} \FF
m u^2.
\end{equation}
In \eqref{prec} we have neglected terms independent of $u$,
originated by the expansion of $k^2$ inside $[\cdots]$
in \eqref{essione},  which have no effect on the
precession of the pericenter.

If
we neglect the last addendum on the right hand side of \eqref{prec},
and we set
\begin{equation}\label{eq:ciai}
c_1 \equiv \frac{m}{h^2}, \qquad
\delta_1 \equiv \frac{3}{2} \FF m,
\qquad
\delta_2 \equiv \frac{A}{h^2} m^2,
\end{equation}
the solution $u_0$ of the corresponding linear equation is
$$
u_0(\phi) = \frac{c_1}{1-\delta_2}
\left\{1+ e \cos \left[\sqrt{1-\delta_2}~ (\phi-\phi_0) \right]
\right\},
$$
where $e$ and $\phi_0$ are  constants of integration. If we neglect $\delta_2$
with respect to $1$, the above solution gives an elliptical
orbit with eccentricity $e$ and semi-latus rectum $p = a (1-e^2) =
\frac{1-\delta_2}{c_1}$.
Substituting $u_0^2$ in place of $u^2$ in \eqref{prec}, the solution $u$
of the corresponding linearized equation is $u = u_0 + u_1$, where
$$
\begin{aligned}
u_1(\phi) = &
\frac{ \delta_1 c_1^2 }{ (1-\delta_2)^2 }
\Big[
\frac{1+e^2/2}{1-\delta_2} +
\frac{e}{\sqrt{1-\delta_2}} (\phi-\phi_0) \sin
\left(
\sqrt{1-\delta_2} ~(\phi-\phi_0)
\right)
\\
& -
\frac{1}{1-\delta_2} \frac{e^2}{6}
\cos\left(
2 \sqrt{1-\delta_2} ~(\phi-\phi_0)
\right)
\Big].
\end{aligned}
$$
If the eccentricity of the orbit is small (i.e., $e^2 << e$)
and we neglect the (small) additive constant
$\frac{\delta_1 c_1^2}{(1-\delta_2)^2} \frac{1+e^2/2}{1-\delta_2}$
which does not contribute
to the precession of the pericenter, we have
$$
u(\phi) \simeq
\frac{c_1}{1-\delta_2} \left\{
1 + e \left[
\cos\left(
\sqrt{1-\delta_2} ~(\phi-\phi_0)
\right)
+
\frac{c_1 \delta_1}{(1-\delta_2)^{3/2}} (\phi-\phi_0)
\sin\left(
\sqrt{1-\delta_2}~(\phi-\phi_0)
\right)
\right]
\right\}.
$$
Taking into account that $\frac{c_1 \delta_1}{(1 -\delta_2)^{3/2}}
(\phi-\phi_0)$
is of order $\em$ and that $\delta_2$ is of order $\em$, we find
\begin{eqnarray*}
u(\phi) && \simeq
\frac{c_1}{1 - \delta_2}
\left[
1 + e \cos \left(
\phi - \phi_0 - c_1 \delta_1 (\phi-\phi_0) - \frac{\delta_2}{2} \left(
1 + 3 c_1 \delta_1
\right) (\phi-\phi_0)
\right)
\right]
\\
&& \simeq
\frac{c_1}{1 - \delta_2}
\left\{
1 + e \cos \left[
(\phi-\phi_0) \left(
1 - c_1 \delta_1 - \frac{\delta_2}{2}
\right)
\right]
\right\}.
\end{eqnarray*}
This is the equation of an elliptic orbit whose pericenter precedes
according to
\begin{equation}\label{veltroni}
(\inttempo \widetilde \omega)_{\rm sec} = 2 \pi
\left(
\frac{1}{1 - c_1 \delta_1 - \frac{\delta_2}{2}}
-1
\right) \simeq 2 \pi \left(
c_1 \delta_1 + \frac{\delta_2}{2}
\right).
\end{equation}
Substituting \eqref{eq:A} and \eqref{eq:ciai} into \eqref{veltroni}
we obtain formula \eqref{erroreguth}.

\section{Acknowledgements}\label{sec:acknowledge}
We thank the University of Roma ``Tor Vergata'', CNR and INFN for
supporting this work.
We wish to thank
B. Bertotti for some
useful advices.



\begin{thebibliography}{99}


\bibitem{Will} C. M. Will, gr-qc/0510072.

\bibitem{shapiro-1999} I.I. Shapiro, Rev. Mod. Phys.  {\bf 71}, Centenary, S41 (1999).

\bibitem{shapiro-mercury} I.I. Shapiro, {\it Gravitation and Relativity 1989},
edited by N. Ashby, D.F. Bartlett, and W. Wyss, Cambridge
University Press, Cambridge, England, 1990, p. 313.

\bibitem{Vessot} R. F. C. Vessot et al., Phys. Rev. Lett. {\bf 45}, 2081 (1980).

\bibitem{VLBI} S.S. Shapiro, J.L. Davis, D.E. Lebach,  J.S. Gregory,
Phys. Rev. Lett. {\bf 92}, 121101 (2004).

\bibitem{reasenberg-shapiro} R.D. Reasenberg et al.,
Astrophys. J. Lett. {\bf 234}, L219–L221 (1979).

\bibitem{Cassini} B. Bertotti, L. Iess, P. Tortora, Nature {\bf 425},
 374 (2003).

\bibitem{Williams2004} J. G. Williams, S. G. Turyshev, D. H. Boggs,
Phys. Rev. Lett. {\bf 93}, 261101 (2004).

\bibitem{CiPa:04} I. Ciufolini, E.C. Pavlis,
Nature {\bf 431}, 958 (2004).

\bibitem{dS:16-2} W. de Sitter,
Monthly Notices of Royal Astr. Soc. {\bf 77}
(Second Paper), 155 (1916).

\bibitem{shap1988} I.I. Shapiro, R.D. Reasenberg, J.F. Chandier,
R.W. Babcock,
Phys. Rev. Lett. {\bf 61}, 2643 (1988).

\bibitem{LT} R. March, G. Bellettini, R. Tauraso, S. Dell'Agnello,
arXiv gr-qc/1101.2791.

\bibitem{Tom} J. Battat et al.,
PASP {\bf 121}, 29 (2009).

\bibitem{He:76} F.W. Hehl, P. von der Heyde, G.D. Kerlick,
J.M. Nester,
Rev. Mod. Phys. {\bf 48}, 393 (1976).

\bibitem{Ha:02}
R.T. Hammond,
Rep. Prog. Phys. {\bf 65},  599
(2002).

\bibitem{StYa:79}
W.R. Stoeger, P.B. Yasskin,
Gen. Rel. Gravit. {\bf 11}, 427 (1979).

\bibitem{YaSt:80}
P.B. Yasskin, W.R. Stoeger,
Phys. Rev. D {\bf 21}, 2081 (1980).

\bibitem{Wei:10}
W-T Ni, Rep. Prog. Phys. {\bf 73}, 056901 (2010).

\bibitem{MaTeGuCa:07}
Y. Mao, M. Tegmark, A.H. Guth, S. Cabi,
Phys. Rev. D {\bf 76}, 1550  (2007).

\bibitem{HeOb:07}
F.W. Hehl, Y.N. Obukhov, Annal. Fondation Louis de Broglie {\bf 32}, 157 (2007).

\bibitem{HaSh:79} K. Hayashi, T. Shirafuji,
Phys. Rev. D {\bf 19}, 3524 (1979).

\bibitem{Ko:82} W. Kopczynski,
J. Phys. A: Math. Gen. {\bf 15}, 493 (1982).

\bibitem{MuNi:83}
F. Mueller-Hoissen, J. Nitsch,
Phys. Rev. D {\bf 18}, 718 (1983).

\bibitem{Ne:88}
J. Nester,
Class. Quant. Grav. {\bf 5}, 1003 (1988).

\bibitem{BlNi:00}
M. Blagojevich, I.A. Nikolic,
Phys. Rev. D {\bf 62}, 024021 (2000).

\bibitem{ArPe:04} H.I. Arcos, J.G. Pereira,
 Int. J. Mod. Phys. D {\bf13},  2193 (2004).

\bibitem{FlRo:07} E.F. Flanagan, E. Rosenthal,
Phys. Rev. D {\bf 75}, 124016 (2007).

\bibitem{PuOb:08}
D. Puetzfeld, Y.N. Obukhov, Phys. Lett. A {\bf 372}, 6711 (2008).

\bibitem{Wi:93} C.M. Will,
{\it Theory and Experiment in Gravitational
Physics}, Cambridge Univ. Press (1993).

\bibitem{CiWh:95}
I. Ciufolini, J.A. Wheeler,
{\it Gravitation and Inertia}, Princeton Univ. Press, Princeton (1995).

\bibitem{Pa:51} A. Papapetrou,
Proc. Roy. Soc. A {\bf 209}, 248 (1951).

\bibitem{BaFr:10} O.V. Babourova, B.N. Frolov,
Phys. Rev. D {\bf 82}, 27503 (2010).

\bibitem{KlPe:99} H. Kleinert, A. Pelster,
Gen. Rel. Grav. {\bf 31}, 1439 (1999).

\bibitem{KlSh:98} H. Kleinert, S.V. Shabanov,
Phys. Lett. B {\bf 428}, 315 (1998).

\bibitem{Kl:00} H. Kleinert,
Gen. Rel. Grav.
{\bf 32}, 769 (2000).

\bibitem{DeTu:02} T. Dereli,
R.W. Tucker, Mod. Phys. Lett. A {\bf 17}, 421 (2002).

\bibitem{CeDeTu:04} H. Cebeci, T. Dereli, R.W. Tucker,
Int. J. Mod. Phys. D {\bf 13}, 137 (2004).

\bibitem{DeTu:82} T. Dereli,
R.W. Tucker, Phys. Lett. B {\bf 110}, 206 (1982).

\bibitem{DeTu:04a} T. Dereli, R.W. Tucker,
arXiv gr-qc/0107017.

\bibitem{BuDeTu:04} D.A. Burton, T. Dereli, R.W. Tucker,
in {\it Symmetries in Gravity and Field Theory}, edited by
V. Aldaya, J.M. Cerver\'o and Y.P. Garcia (Ediciones Universidad
Salamanca, 2004),
p. 237.

\bibitem{Po:71} V.N. Ponomariev,
Bull. Acad. Polon. Sci. {\bf XIX}, 6 (1971).

\bibitem{BrCl:61} D. Brouwer, G.M. Clemence, {\it Methods of Celestial
Mechanics}, Academic Press (1961).

\bibitem{GeWe:71} F.T. Geyling, H.R. Westerman,
{\it Introduction to Orbital Mechanics},
Addison Wesley (1971).

\bibitem{dS:proc} W. de Sitter,
KNAW, Proceedings, {\bf 19} I, Amsterdam, 367 (1917).

\bibitem{FiKr:76} A.M. Finkelstein, V. Ja. Kreinovich,
Celestial Mech. {\bf 13}, 151 (1976).

\bibitem{KoRuTa} V.A. Kostelecky, N. Russell, J. Tasson,
   Phys. Rev. Lett.  {\bf 100}, 111102 (2008).

\bibitem{HeAd}
B.R. Heckel et al.,
   Phys. Rev.  D {\bf 78}, 092006 (2008).

\bibitem{nordtvedt-effect} K. Nordtvedt, Jr., Phys. Rev. {\bf 169}, 1014 (1968);
K. Nordtvedt, Jr., Phys. Rev. {\bf 169}, 1017 (1968);
K. Nordtvedt, Jr., Phys. Rev. {\bf 170}, 1186 (1968).

\bibitem{To:91} N. Toma,
Progr. Theor. Phys. {\bf 86}, 659 (1991).

\bibitem{SBI} M. Fermi et al.
 ``37th COSPAR Scientific Assembly'' (2008), July 13-20,
Montr\'eal, Canada,  868.

\bibitem{Bender} P. L. Bender, Adv. Space Res. {\bf 14}, 233-242 (1994).

\bibitem{IAC} D. G. Currie et al., ``60th International Astronautical Congress'',
Daejeon, Korea, October 12-16, 2009, Paper n. IAC-09.A2.1.11.

\bibitem{ILRS08} S. Dell'Agnello et al., in ``Proceedings of the 16th
International Workshop on Laser Ranging'' (2008), October 13-17, Poznan, Poland, 121.

\bibitem{INFNPaper:10} S. Dell'Agnello et al.,
Adv. Space Res., {\it Galileo Special Issue}, {\bf 47} (2011) 822-842.

\bibitem{MAGIA} S. Dell'Agnello et al.,
Exp. Astron., {\it MAGIA Special Issue}, DOI 10.1007/s10686-010-9195-0 (2010).

\bibitem{Milani} A. Milani et al., Phys. Rev. D {\bf 66}, 082001 (2002).

\bibitem{Ashby} N. Ashby, P. L. Bender, J. M. Warh, Phys. Rev. D {\bf 75}, 022001 (2007).

\bibitem{Slava} S. G. Turyshev, J. D. Anderson, R. W. Hellings,
 arXiv:gr-qc/9606028v1.

 \bibitem{Slava1}
J. D. Anderson et al.,
Planetary Space Sci.  {\bf 45}, 21 (1997).

\bibitem{MaTeGuCa:07online}
Y. Mao, M. Tegmark, A.H. Guth, S. Cabi,
arXiv:gr-qc/0608121v4.

\end{thebibliography}
\end{document}